\documentclass[aps,onecolumn,preprint,pre,superscriptaddress]{revtex4-1}

\usepackage{graphicx}
\usepackage{amsmath}
\usepackage{amssymb}
\usepackage{color}

\begin{document}
	
\title{Accuracy of position determination in Ca$^{2+}$ signaling}

\author{Vaibhav H. Wasnik}
\affiliation{NCCR Chemical Biology, Departments of Biochemistry and Theoretical Physics, University of Geneva, 1211 Geneva, Switzerland}
\affiliation{Indian Institute of Technology Goa, Ponda-403401, India}
	
\author{Peter Lipp}
\affiliation{Institute for Molecular Cell Biology, Research Centre for Molecular Imaging and Screening, Center for Molecular Signaling (PZMS), Medical Faculty, Saarland University, Homburg/Saar, Germany}

\author{Karsten Kruse}
\affiliation{NCCR Chemical Biology, Departments of Biochemistry and Theoretical Physics, University of Geneva, 1211 Geneva, Switzerland}

\date{\today}

\begin{abstract} 
A living cell senses its environment and responds to external signals. In this work, we study theoretically, the precision at which cells can determine the position of a spatially localized transient extracellular signal. To this end, we focus on the case, where the stimulus is converted into the release of a small molecule that acts as a second messenger, for example, Ca$^{2+}$, and activates kinases that change the activity of enzymes by phosphorylating them. We analyze the spatial distribution of phosphorylation events using stochastic simulations as well as a mean-field approach. Kinases that need to bind to the cell membrane for getting activated provide more accurate estimates than cytosolic kinases. \textcolor{black}{Our results could explain why the rate of Ca$^{2+}$ detachment from the membrane-binding conventional Protein Kinase C$\alpha$ is larger than its phosphorylation rate.}
\end{abstract}

	
\maketitle

\section{Introduction}

Living cells respond to external stimuli. Often these stimuli consist in binding a ligand to a cell surface receptor~\cite{Alberts2008}, but cells also react to mechanical forces that are applied to them~\cite{Janmey:2007el}, to changes in ambient light or temperature. In this way, cells can migrate towards favourable and away from unfavourable environmental conditions~\cite{Swaney:2010by}, sense the density of cells of their own kind~\cite{Ng:2009fm}, initiate or inhibit cell division, or trigger developmental programs~\cite{Engler:2006ga}. The ability to sense external stimuli is thus of paramount importance for the success of a single- or multi-celled species to survive: The better a cell can read out signals, the better it will do.

This has led to investigations of the physical limits of cellular signalling. Berg and Purcell determined the conditions under which an organism can optimally determine the concentration of a molecule in its environment in case the cell uses independent receptors~\cite{BERG:1977uj}. In the bacterium \textit{Escherichia coli} some kind of chemoreceptor clusters at one cell end~\cite{Maddock:1993tu}, and physical studies revealed how this clustering can enhance sensitivity to external signals~\cite{Bray:1998vj,Duke:1999td,Mello:2003cz,Mello:2004ed,Mello:2005fm,Endres:2006ix}. Subsequently, these investigations have been generalized to account also for the energetics of ligand-receptor binding~\cite{Bialek:2005kf,Bialek:2008hb}. Whereas \textit{E.~coli} is too small to directly sense spatial gradients in the concentration of chemoattractants and thus relies to this end on detecting temporal concentration changes while moving, eukaryotic cells like neutrophils and leukocytes of the human immune system, the budding yeast \textit{Saccharomyces cerevisiae}, or the slime mold \textit{Dictyostelium discoideum} can directly sense spatial gradients~\cite{VanHaastert:2004bs}. Physical constraints on the accuracy of sensing spatial gradients by single cells have been established~\cite{Endres:2008eb,Rappel:2008jv,Rappel:2008bh}. In this context, spatial aspects of the intracellular signalling network have been considered~\cite{Xiong:2010fb}.

In addition to sensing spatial gradients, eukaryotic cells can also respond to spatially localized signals. For example, neurons reinforce or weaken synapses in response to transient stimuli~\cite{Deisseroth:1996ht,Wheeler:2008ja}. Another example is provided by cells from the immune system, which polarize upon making contact with an antigen presenting cell~\cite{Kapsenberg:2003it}.

In this work, we address the question of how the spatial resolution at which a cell can detect a stimulus depends on the intracellular mechanism of reading it out. Stimuli of any kind are typically transformed into a cell response by activating or inactivating proteins through adding or removing phosphate groups (phosphorylation and dephosphorylation) to and from certain amino acids. Proteins leading to phosphorylation are called kinases, whereas phosphatases carry out dephosphorylation. Eventually, changing the phosphorylation of proteins can lead to changes of the cytoskeletal organization on short time scales and to modifications of the cellular gene expression profile on long time scales to give but two examples. Commonly, external stimuli do not directly (de-)phosphorylate proteins, but first elicit the release of a second messenger, for example, cyclic Adenosine monophosphate (cAMP), inositol triphosphate (IP$_3$), diacylglycerol (DAG), or Ca$^{2+}$. In this way, the same machinery can be used to respond to a variety of different stimuli. At the same time it raises the question of how to obtain a specific response to a given signal. 

The second messenger Ca$^{2+}$ is read out by different proteins. Notably, a number of kinases are activated by Calmodulin, a peptide diffusing in the cytoplasm that changes conformation after binding to a Ca$^{2+}$ ion. In contrast, the ubiquitously expressed conventional Protein Kinase C$\alpha$ (PKC$\alpha$) directly binds Ca$^{2+}$. For activation it requires also binding to DAG in the plasma membrane of a cell. Strikingly, the detachment rate of Ca$^{2+}$ from PKC$\alpha$ has been measured to be of the order of 50ms, whereas the phosphorylation of a target protein by PKC$\alpha$ takes about 500ms~\cite{Nalefski:2001wy}. This suggests that PKC$\alpha$ is an inefficient kinase and raises the question of the evolutionary benefit of a high Ca$^{2+}$ detachment rate.

In this work, we develop a framework for studying the spatial distribution of phosphorylation events inside a cell. We start by studying a toy model that allows us to introduce some notation and to present the tools we will use for analyzing the stochastic processes corresponding to phosphorylation by a cytosolic and by a membrane-\textcolor{black}{binding} kinase. We will then study the spatial distribution of phosphorylation events in response to a single Ca$^{2+}$ for a cytosolic and a membrane-\textcolor{black}{binding} kinase, where we find that \textcolor{black}{the latter} yields a more accurate read-out than \textcolor{black}{the former}. We will then discuss responses to Ca$^{2+}$ puffs and investigate the influence of background phosphorylation on the precision of estimating the site of Ca$^{2+}$ entry. A short account of some of the results presented in this work has been given in Ref.~[Letter].

\section{A toy model}

\begin{figure}
\includegraphics[width=0.5\textwidth]{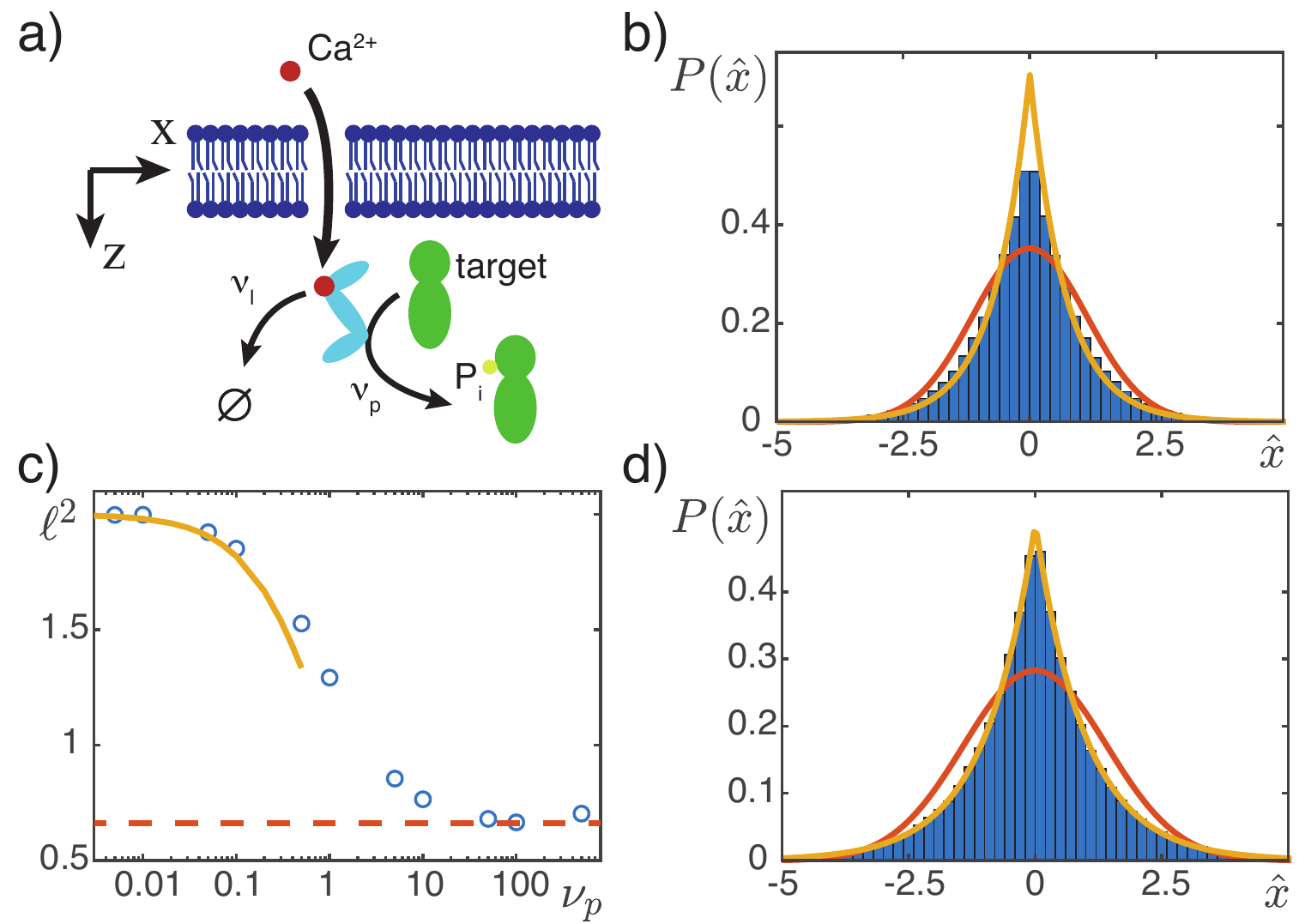}
\caption{\label{fig:toyModel}(color online) Toy model for position determination of a Ca$^{2+}$ entry site. a) Illustration of the system. The Ca$^{2+}$ ion enters through the membrane at $x=0$ and immediately activates a kinase, which phosphorylates target proteins at rate $\nu_p$. The Ca$^{2+}$ ion is lost immediately after detaching from the kinase at rate $\nu_l$. b) Distribution $P$ of estimated Ca$^{2+}$ entry sites $\hat{x}$ obtained from $10^6$ stochastic simulations for $\nu_p=1$. c) Estimation error $\ell^2$ as a function of the phosphorylation rate $\nu_p$. Circles are from stochastic simulations. The yellow full line represents the error for $\nu_p\ll1$, Eq.~(\ref{eq:errorToyModelNupSmall}), the red dashed line the error for $\nu_p\to \infty$, Eq.~(\ref{eq:errorToyModelNupLarge}). d) Distribution $P$ of estimated Ca$^{2+}$ entry sites $\hat{x}$ obtained from $10^7$ stochastic simulations for $\nu_p=0.01$. In (b, d), the red line presents a Gaussian fit to the data, whereas the yellow line is the normalized distribution $P$ for $\nu_p\ll1$ given in Eq.~(\ref{eq:distributionToyModelNupSmall}). }
\end{figure}

In order to introduce some quantities as well as some methods that we will use later to analyze the localization of Ca$^{2+}$ influx into the system, we will study a toy model in this section. It can be interpreted as describing kinase-dependent phosphorylation following the entry of a single Ca$^{2+}$ ion, see Fig.~\ref{fig:toyModel}a. The kinase is activated by immediately binding the Ca$^{2+}$ ion at the latter's entry site. After inactivation of the kinase following Ca$^{2+}$ detachment, the Ca$^{2+}$ ion is immediately lost.

\subsection{The Master equation}

Consider activation of the kinase at $x=0$. We will restrict attention to the dynamics along the $x$-direction and assume that the system has no boundaries in that direction. The activated kinase phosphorylates at a constant rate $\nu_p$, while diffusing with a diffusion constant $D$. The kinase is inactivated at rate $\nu_l$. Let $n$ denote the spatial distribution of phosphorylation events. The state of the system is given by the probability $\mathcal{P}_a\left[n(\xi);x,t\right]$ for having an active kinase at $x$ at time $t$ and a distribution of phosphorylation events $n(\xi)$. Similarly, $\mathcal{P}_i\left[n(\xi);t\right]$ is the corresponding distribution when the kinase is inactive. In this case its position is irrelevant\textcolor{black}{, because the kinase cannot be activated again such that no further phosphorylation events can cccur}. The Master equation governing the time evolution of these distributions is given by
\begin{align}
\label{eq:masterEquationToyModelInactive}
\partial_t\mathcal{P}_i\left[n(\xi);t\right] &= \nu_l\int \mathrm{d}x\;\mathcal{P}_a\left[n(\xi);x,t\right]\\
\label{eq:masterEquationToyModelActive}
\partial_t\mathcal{P}_a\left[n(\xi);x,t\right] &= D\partial_x^2\mathcal{P}_a\left[n(\xi);x,t\right]-\nu_l\mathcal{P}_a\left[n(\xi);x,t\right]\nonumber\\
&\quad\quad +\nu_p\left\{\mathcal{P}_a\left[n(\xi)-\delta(\xi-x);x,t\right]-\mathcal{P}_a\left[n(\xi);x,t\right]\right\},
\end{align} 
where $\delta$ denotes the Dirac distribution. The initial condition at $t=0$ is given by 
\begin{align}
\mathcal{P}_a\left[n(\xi)=0;x,t=0\right] &=\delta(x)
\end{align}
and all other probabilities equal to zero. 

For the analysis of the Master equation, we will scale time by $\nu_l$ and space by $\sqrt{D/\nu_l}$. The only remaining dimensionless parameter is then $\nu_p/\nu_l$. We will keep the notation $\nu_p$ for the dimensionless phosphorylation rate.

\subsection{Number of phosphorylation events}
It is instructive to first neglect the spatial aspects of the phosphorylation dynamics and to determine the distribution of the number of phosphorylation events. Consider the probability distribution $P_a(n,t)$ that $n$ phosphorylation events have taken place at time $t$ and with the kinase being active. The corresponding distribution when the kinase is inactive is $P_i(n,t)$. The Master equation for $P_a$ and $P_i$ reads
\begin{align}
\label{eq:Pp0dot}
\frac{\mathrm{d}}{\mathrm{d}t}P_i(n) &=  P_a(n)\\
\label{eq:Ppndot}
\frac{\mathrm{d}}{\mathrm{d}t}P_a(0) &= - P_a(0) - \nu_pP_a(0)\\
\label{eq:Pp1dot}
\frac{\mathrm{d}}{\mathrm{d}t}P_a(n) &= - P_a(n) + \nu_p\left(P_a(n-1)-P_a(n)\right),
\end{align}
where $n\ge0$ in Eq.~(\ref{eq:Pp0dot}) and $n\ge1$ in Eq.~(\ref{eq:Pp1dot}). \textcolor{black}{The first equation describes the inactivation of the kinase. In Eqs.~(\ref{eq:Ppndot}) and (\ref{eq:Pp1dot}), the first terms on the left hand side account for inactivation of the kinase, whereas the remaining terms describe phosphorylation.} The initial condition is $P_a(0,t=0)=1$, $P_a(n,t=0)=0$ for all $n>0$, and $P_i(n,t=0)=0$ for all $n\ge0$. The solution to these equations is
\begin{align}
P_i(n,t) &= (1-r)r^n-(1-r)\left[\sum_{k=0}^n \frac{\left(\nu_pt\right)^k}{k!}r^{n-k}\right]\mathrm{e}^{-\left(1+\nu_p\right)t}\\
P_a(n,t)&=\frac{\left(\nu_pt\right)^n}{n!}\mathrm{e}^{-\left(1+\nu_p\right)t}\\
\intertext{with}
r &= \frac{\nu_p}{1+\nu_p}
\end{align}
for all $n\ge0$. For $t\to\infty$ we have
\begin{align}
P_i(n) &= (1-r)r^n\\
P_a(n) &= 0,
\end{align}
which yields the average number $\textcolor{black}{N_\mathrm{p}\equiv}\langle n\rangle=\sum_{n=0}^\infty n\left(P_i(n,t)+P_a(n,t)\right)$ of phosphorylation events, namely $\textcolor{black}{N_\mathrm{p}}=\nu_p$. The variance around this value is $\textcolor{black}{\mathrm{var}(n)\equiv}\langle n^2\rangle-\langle n\rangle^2=\nu_p\left(1+\nu_p\right)=(1+\textcolor{black}{N_\mathrm{p}})\textcolor{black}{N_\mathrm{p}}$.

\subsection{Estimated position of Ca$^{2+}$ release and estimation error}
For a given distribution of phosphorylation events $n(\xi)$, we estimate the position of Ca$^{2+}$ release by computing the distribution average:
\begin{align}
\label{eq:estimatedAverage}
\hat{\xi}_{n(\xi)} &= \int\mathrm{d}\xi\;\xi n(\xi)/\int\mathrm{d}\xi\;n(\xi),
\end{align}
which obviously only exists, if $\int\mathrm{d}\xi\;n(\xi)\neq0$. If $n(\xi)= 0$ for all $\xi$, the position cannot be estimated. In this way, we obtain a distribution $P$ of estimated positions. Explicitly, this distribution is given by 
\begin{align}
P(\hat{\xi},t) &= \int\mathcal{D}n(\xi)\;\delta(\hat{\xi}_{n(\xi)}-\hat{\xi})\mathcal{P}[n(\xi);t],
\end{align}
where $\hat{\xi}_{n(\xi)}$ is the average of the distribution $n(\xi)$ according to Eq.~(\ref{eq:estimatedAverage}) and $\mathcal{P}[n(\xi);t]\equiv\mathcal{P}_i[n(\xi);t] + \int\mathrm{d}x\;\mathcal{P}_a[n(\xi);x,t]$ is the probability functional of phosphorylation distributions. We will use the variance $\ell^2$ of the distribution $P$ as a measure of the error in localizing the site of Ca$^{2+}$ release. Given that the mean of the estimated positions is zero, we obtain
\begin{align}
\label{eq:estimationError}
\ell^2 &= \int\mathrm{d}\hat{\xi}\;\hat{\xi}^2P(\hat{\xi},t).
\end{align}
Combining Eqs.~(\ref{eq:estimatedAverage})-(\ref{eq:estimationError}), we get
\begin{align}
\label{eq:estimationError2}
\ell^2 &= \int\mathcal{D}n(\xi)\;\left(\frac{\int \mathrm{d}\xi\;\xi n(\xi)}{\int \mathrm{d}\xi\;n(\xi)}\right)^2 \mathcal{P}[n(\xi);t].
\end{align}

\subsection{Solution of the Master equation}
\label{sec:numericalSolution}

In general, it is not possible to solve the Master equation~(\ref{eq:masterEquationToyModelInactive}) and (\ref{eq:masterEquationToyModelActive}) and to determine the estimation error analytically. Let us thus start our analysis by a stochastic simulation. To this end, we use a variant of the Gillespie algorithm: we define the total rate of possible events as $\nu_\mathrm{tot}\equiv1+\nu_p$ and draw the time $\Delta t$ until occurrence of the next event from the distribution $\nu_\mathrm{tot}\exp\left(-\nu_\mathrm{tot}t\right)$. Then, the kind of event is determined by drawing a number from a uniform distribution on the interval $[0,1+\nu_p]$. If the number is smaller than $1$, then the kinase is inactivated and the simulation stops. In the opposite case, the position of the kinase is changed by drawing a random number from a Gaussian distribution with zero mean and variance $2\Delta t$ and adding this value to the current position of the kinase. This position is then recorded as the site of phosphorylation and the simulation continues with drawing the time to the next event. From the distribution of positions of phosphorylation events obtained in this way, we calculate the mean position, which we take to be the estimated position at which the Ca$^{2+}$ ion entered the system. If there was no phosphorylation event, then the position of Ca$^{2+}$ entry is not estimated.

In Figure~\ref{fig:toyModel}b, we show the distribution of estimated Ca$^{2+}$ entry sites for $\nu_p=1$ obtained form $10^6$ simulation runs. The dependence of the error $\ell^2$ on the phosphorylation rate $\nu_p$ is given in Fig.~\ref{fig:toyModel}c, where for each value at least $10^6$ simulation runs have been performed. The error is about 2 for $\nu_p\to 0$ and decreases with increasing phosphorylation rate. This is expected as for increasing values of $\nu_p$ an increasing number of phosphorylations and thus position measurements occur on average for a Ca$^{2+}$ ion. In the limit $\nu_p\to\infty$, the error is $\ell^2\approx\frac{2}{3}$.

We can make some analytical progress by considering the limiting cases of very small and large phosphorylation rates.

\subsubsection{The limit $\nu_p\ll1$}
For very small phosphorylation rates $\nu_p\ll1$, the probability of having a trajectory with two or more phosphorylation events is negligible. In that case, the Master equation~(\ref{eq:masterEquationToyModelInactive}) and (\ref{eq:masterEquationToyModelActive}) reduces to
\begin{align}
\label{eq:pa0NupEq0}
\partial_t P_a^0 &= \partial_x^2 P_a^0 -\left(1+\nu_p\right)P_a^0\\
\label{eq:pa1NupEq0}
\partial_t P_a^1 &= \partial_x^2 P_a^1 + \nu_p\delta(x-\xi) P_a^0 - P_a^1\\
\partial_t P_i^1 &= \int\mathrm{d}x\;P_a^1 \equiv \bar{P}_a^1,
\end{align}
where $P_a^0(x,t)$ is the probability of having, at time $t$, an active kinase at $x$ without any phosphorylation, whereas $P_a^1(\xi;x,t)$ is the probability of the active kinase being at $x$ and where a phosphorylation event had occurred at position $\xi$. Similarly, $P_i^1(\xi;t)$ denotes the corresponding probability after inactivation of the kinase. We do not show the dynamic equation for the probability of having lost the kinase before it phosphorylated, because it is irrelevant for estimating the position of the Ca$^{2+}$ release site. In the last equation, we have introduced the marginal distribution of the phosphorylation position after integrating out the position of the kinase, $\bar{P}_a^1(\xi,t)=\int\mathrm{d}x\;P_a^1(\xi;x,t)$. Integrating Eq.~(\ref{eq:pa1NupEq0}) with respect to the kinase position $x$, we obtain its dynamic equation
\begin{align}
\partial_t \bar{P}_a^1(\xi,t) &= \nu_p P_a^0(\xi,t) - \bar{P}_a^1(\xi,t).
\end{align}

The solution to Eq.~(\ref{eq:pa0NupEq0}) is
\begin{align}
P_a^0\left(x,t\right) &=\frac{1}{\sqrt{4\pi t}}\exp\left\{-\left(1+\nu_p\right)t-\frac{x^2}{4t}\right\}
\end{align}
The distribution of the phosphorylation position $P(\xi)\equiv\bar{P}_a^1(\xi)+P_i^1(\xi)$ is obtained from the expression for $P_a^0$ through $P(\xi,t)=\nu_p\int_0^t\mathrm{d}t^\prime P_a^0\left(\xi,t^\prime\right)$. In the limit $t\to\infty$, we get
\begin{align}
\label{eq:distributionToyModelNupSmall}
P(\xi) &= \frac{\nu_p}{2\sqrt{1+\nu_p}}\exp\left\{-\sqrt{1+\nu_p}\left|\xi\right|\right\},\\
\intertext{such that}
\label{eq:errorToyModelNupSmall}
\ell^2&=\frac{2}{1+\nu_p}.
\end{align}
For small enough values of $\nu_p$, the distribution Eq.~(\ref{eq:distributionToyModelNupSmall}) agrees well with the distribution of estimated entry sites obtained from stochastic simulations, see Fig.~\ref{fig:toyModel}d. The expression of the error as given by Eq.~(\ref{eq:errorToyModelNupSmall}) gives a good approximation for $\nu_p\lesssim0.1$.

\subsubsection{Continuous phosphorylation}
We now consider the limit of a very large phosphorylation rate, such that phosphorylation occurs at any point of the kinase's trajectory. In that case, the estimated position of release $\hat{x}$ of a single Ca$^{2+}$ as defined in Eq.~(\ref{eq:estimatedAverage}) is obtained by the average position of the kinase along its trajectory $x(t)$, that is, 
\begin{align}
\hat{x}_T &=\frac{1}{T}\int_0^T\mathrm{d}t\; x(t),
\end{align}
where $T$ is the time span between entry of the Ca$^{2+}$ and loss of the kinase. The error is then given by the variance
\begin{align}
\label{eq:errorContinuousPhosphorylation}
\ell^2 &= \int_0^\infty\mathrm{d}T\;\left\langle \hat{x}_T^2\right\rangle \mathrm{e}^{-T},
\end{align}
where the exponential factor accounts for the probability of finding a trajectory of duration $T$.

To calculate the the expectation value $\left\langle \hat{x}_T^2\right\rangle$, we note that the trajectories $x(t)$ are solutions to a Langevin equation 
\begin{align}
\dot x(t)&=\zeta(t)
\end{align}
for $0\le t\le T$, where $\zeta$ is a fluctuating ``force'' that obeys a Gaussian distribution at each time $t$ with zero mean, $\left\langle \zeta(t)\right\rangle=0$, and $\left\langle \zeta(t)\zeta(t^\prime)\right\rangle=2\delta(t-t^\prime)$. The solution to this equation is $x(t)=\int_0^t\mathrm{d}t^\prime\;\zeta\left(t^\prime\right)$, such that
\begin{align}
\left\langle \hat{x}_T^2\right\rangle &= \frac{1}{T^2}\int_0^T\mathrm{d}t\int_0^T\mathrm{d}\bar{t}\left\langle x(t)x(\bar{t})\right\rangle\\
&=\frac{1}{T^2}\int_0^T\mathrm{d}t\int_0^T\mathrm{d}\bar{t}\int_0^t \mathrm{d}t^\prime\int_0^{\bar{t}} \mathrm{d}t^{\prime\prime}\left\langle \zeta(t^\prime)\zeta(t^{\prime\prime})\right\rangle\\
&=\frac{1}{T^2}\int_0^T\mathrm{d}t\int_0^T\mathrm{d}\bar{t}\;2\left|t-\bar{t}\right|\\
&=\frac{2}{3}T.
\end{align}
From this we obtain using Eq.~(\ref{eq:errorContinuousPhosphorylation})
\begin{align}
\label{eq:errorToyModelNupLarge}
\ell^2 &= \frac{2}{3},
\end{align}
which is in good agreement with the numerical results presented in Fig.~\ref{fig:toyModel}c.

\subsection{The mean-field limit}
\label{sec:meanfieldToyModel}
A natural approximation is to use a mean-field \textit{ansatz} and to pose that the rate of phosphorylation at position $x$ is proportional to the probability of finding the kinase at this position. Let $p_a$ denote the probability of finding an active kinase at $x$, $p_a(x,t)=\int\mathcal{D}n(\xi)\;\mathcal{P}_a\left[n(\xi);x,t\right]$, where the functional integral extends over all possible phosphorylation distributions $n(\xi)$ with $n(\xi)\ge0$ for all $\xi$. Furthermore, let $p_i$ denote the probability of having no kinase, $p_i(t)=\int\mathcal{D}n(\xi)\;\mathcal{P}_i\left[n(\xi);t\right]$. The time-evolution of $p_a$ obeys
\begin{align}
\partial_t p_a&= \partial_x^2 p_a - p_a.
\end{align}
Once $p_a$ is known, then $p_i = 1-\int\mathrm{d}x\; p_a(x)$. Furthermore, the probability of having $n$ phosphorylation events at position $x$ at time $t$, $P(n,x,t)$, is determined by 
\begin{align}
\dot{P}(0,x,t) &=-\nu_p p_a(x,t) P(0,x,t)\\
\dot{P}(n,x,t) &= \nu_p p_a(x,t) \left(P(n-1,x,t)-P(n,x,t)\right)
\end{align}
for $n\ge1$. The normalization conditions read $\sum_{n=0}^\infty P(n,x,t)=1$ for all $x$ and all $t$ and $\int_{-\infty}^\infty \mathrm{d}x\;p_a(x,t)+p_i(t)=1$ for all $t$. The initial condition is $P(0,x,t=0)=1$ for all $x$ and $p_a(x,t=0)=\delta(x)$. 

The solution to these equations is
\begin{align}
\label{eq:probabilityMF}
P(n,x,t) &= \frac{\nu_p^n}{n!}\bar p_a(x,t)^n\exp\left\{-\nu_p\bar p_a(x,t)\right\}\\
\intertext{with}
\bar p_a(x,t) &=\int_0^t\mathrm{d}t^\prime p_a(x,t^\prime)\\
\intertext{and}
p_a(x,t) &=\sqrt{\frac{\pi}{2t}}\exp\left\{ -t- \frac{x^2}{4t}\right\}.
\end{align}
We are interested in the distribution as $t\to\infty$. In that case, 
\begin{align}
\bar p_a(x) &=  \frac{1}{2}\exp\{-|x|\},
\end{align}
where the bar indicates the distribution for $t\to\infty$.

In the spirit of the mean-field \textit{ansatz}, we replace the expression for the estimation error, see Eq.~(\ref{eq:estimationError2}), by
\begin{align}
\label{eq:errorMF}
\ell^2 &=\frac{\int\mathrm{d}x\;x^2\hat{n}(x)}{\int\mathrm{d}x\;\hat{n}(x)}.
\end{align}
Here, $\hat{n}$ is the mean number of phosphorylation events at $x$. As we will see below in Sect.~\ref{sec:puffMF}, the expression for the estimation error as defined in Eq.~(\ref{eq:estimationError2}) differs from the above expression if the probability distribution is given by Eq.~(\ref{eq:probabilityMF}). However, the two expressions for the error are the same in the limit of a small number of phosphorylation events, $\langle n\rangle\ll1$. 

From the above solution for $P(n,x,t)$ in the limit $t\to\infty$, we get for the mean number of phosphorylation events at $x$ $\hat{n}(x)=\nu_p\bar p_a(x)$ and thus
\begin{align}
\ell^2&=1.
\end{align}
In the rescaled units, this is just the diffusion length of the activated kinase, $\sqrt{D/\nu_l}$ in the original units. Note, that it holds for arbitrary values of $\nu_p>0$, which is different from the results of the stochastic simulations. \textcolor{black}{Indeed, the mean-field equations can only be expected to work well in cases, when the number of phosphorylation events is small, that is, when two or more phosphorylation events for a single Ca$^{2+}$ ion are rare. In fact, if there is only one phosphorylation event, then the spatial distribution of the active kinase determines directly the distribution of phosphorylation events. This is not true for subsequent phosphorylations.} In spite of this obvious failure of the mean-field approximation to determine the dependence of the error on $\nu_p$, it does give, however, the correct order of magnitude of the error, which varies between 2/3 and $2$, see Fig.~\ref{fig:toyModel}d. 

\section{Phosphorylation dynamics in response to a single Ca$^{2+}$ ion}

\begin{figure}
\includegraphics[width=0.5\textwidth]{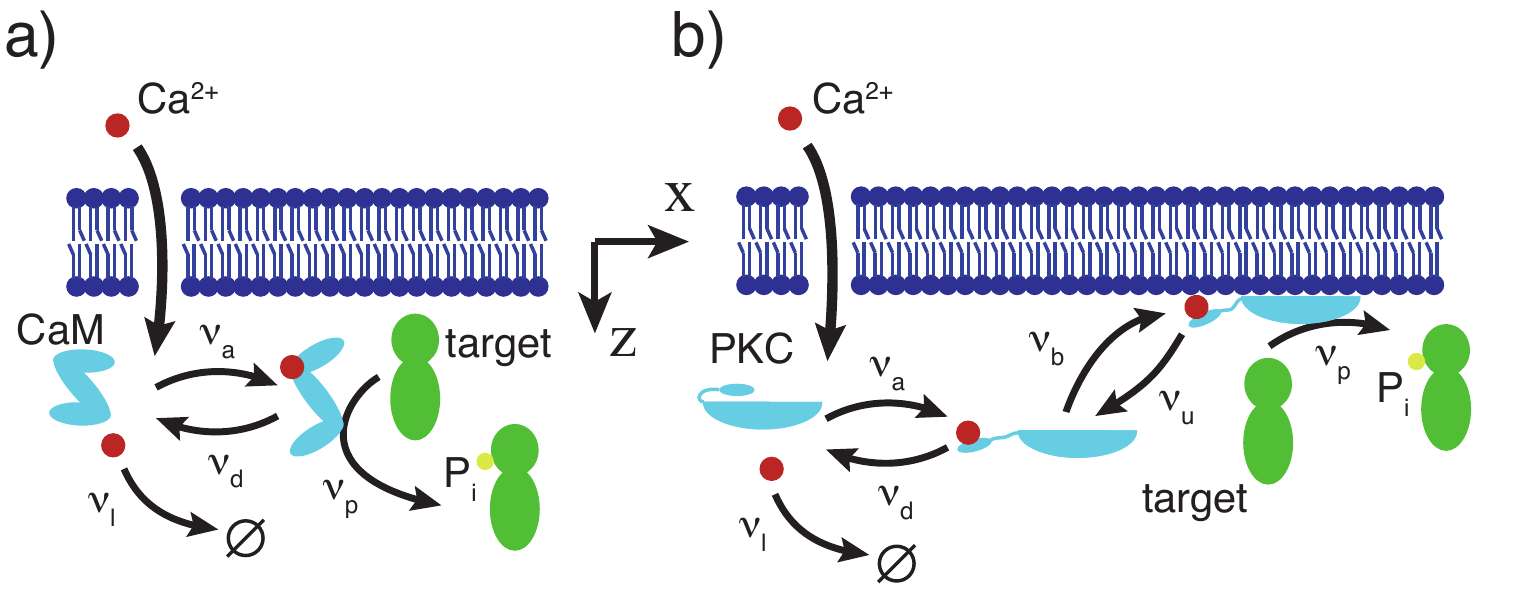}
\caption{\label{fig:illustration}(color online) Schematic of the two scenarios of Ca$^{2+}$ dependent phosphorylation. a) Cytosolic kinase: A cytosolic kinase is activated directly by Ca$^{2+}$. The Ca$^{2+}$ attaches at rate $\nu_a$ and detaches at rate $\nu_d$. The activated kinase phosphorylates target proteins at rate $\nu_p$ and Ca$^{2+}$ is lost at rate $\nu_l$. b) Membrane-binding kinase: In this case, after attaching Ca$^{2+}$, the kinase still needs to bind to the membrane before it is active. Binding and unbinding occur at rates $\nu_b$ and $\nu_u$.}
\end{figure}

We will now study the phosphorylation dynamics and the ensuing spatial distributions of phosphorylation events in two scenarios that reflect essential properties of Ca$^{2+}$ activated kinases in response to a single Ca$^{2+}$ ion entering the system, see Fig.~\ref{fig:illustration}. A single Ca$^{2+}$ ion entering a real cell is unlikely to elicit a response. However, when treating the case of a Ca$^{2+}$ puff below, we will assume that all Ca$^{2+}$ ions are independent of each other. Therefore, studying the response to a single Ca$^{2+}$ ion is appropriate. As we will see in Sect.~\ref{sec:puffs}, however, the passage from one Ca$^{2+}$ ion to a Ca$^{2+}$ puff is not trivial, because not all Ca$^{2+}$ ions lead to a phosphorylation event.

The Ca$^{2+}$ ion enters the system at $x=z=0$ at $t=0$, where the $z$-direction is the direction perpendicular to the membrane, which is located at $z=0$. In the first scenario, the Ca$^{2+}$ ion diffuses in the cytoplasm until it either binds to and thereby activates a kinase or gets lost from the system. The Ca$^{2+}$ ion can detach and reattach to the kinase and can only be lost from the system, when not being attached to the kinase. In the second scenario, after attaching a Ca$^{2+}$ ion, the kinase still needs to bind to the membrane to be activated. 

As in the toy model, we will consider the average position of phosphorylation events in response to a Ca$^{2+}$ ion as an estimate of the Ca$^{2+}$ entry site. This is adequate if the rate of dephosphorylation is much smaller than the rate at which the Ca$^{2+}$ ion is eventually lost form the system. It also requires that the phosphorylated target proteins remain immobile if the estimate obtained in this way should give a good proxy for the localization of the cell response. The dynamics of the target proteins, however, depends on the protein at hand and can meaningfully be studied only with a specific cell response in mind. In contrast, the results we obtain are independent of the specific response and provide a general lower bound for the estimation error. 

\subsection{The Master equations}
\label{sec:masterEquations}

In this section, we present the Master equations for the two different scenarios of phosphorylation dynamics. We will consider only the situation in which kinases are abundant and uniformly distributed, such that the attachment rate of Ca$^{2+}$ to a kinase is constant.

\subsubsection{Cytosolic kinase}

Consider a kinase that is active immediately after attaching a Ca$^{2+}$ ion. In fact, there are no cytosolic kinases known that are directly activated by Ca$^{2+}$ binding. Instead, Ca$^{2+}$ typically binds to Calmodulin (CaM), which then binds to and thereby activates a Calmodulin-dependent kinase~\footnote{There are also Calmodulin phosphatases, which dephosphorylate target proteins. \textit{Cum grano salis} the results developed in this work also apply to phosphatases.}. In our analysis, we neglect the second step. If anything, direct activation by Ca$^{2+}$ will only increase the accuracy of position determination.

For a cytosolic kinase that is directly activated by binding a Ca$^{2+}$ ion, the system state is determined by the spatial distribution of phosphorylation events $n(\xi)$ and, if the Ca$^{2+}$ has not been lost from the system, by its position $x$, where we have to distinguish between the possibility that the Ca$^{2+}$ ion is either free or attached to a kinase. Let $\mathcal{P}_i$ be the corresponding probability distribution, when Ca$^{2+}$ is present, but not attached to the kinase, $\mathcal{P}_a$ the probability distribution in the case that Ca$^{2+}$ is attached to a kinase, and $\mathcal{P}_0$ the probability distribution, when the Ca$^{2+}$ ion is lost from the system. Then the Master equation is given by
\begin{align}
\label{eq:masterEquationDiffusiveKinaseInactive}
\partial_t\mathcal{P}_i\left[n(\xi);x,t\right] &= D_C\partial_x^2\mathcal{P}_i\left[n(\xi);x,t\right]+\nu_d\mathcal{P}_a\left[n(\xi);x,t\right]\nonumber\\
&\quad\quad- (\nu_a+\nu_l)\mathcal{P}_i\left[n(\xi);x,t\right]\\
\label{eq:masterEquationDiffusiveKinaseActive}
\partial_t\mathcal{P}_a\left[n(\xi);x,t\right] &= D_K\partial_x^2\mathcal{P}_a\left[n(\xi);x,t\right]-\nu_d\mathcal{P}_a\left[n(\xi);x,t\right] +
\nu_a\mathcal{P}_i\left[n(\xi);x,t\right]\nonumber\\
&\quad\quad +\nu_p\left\{\mathcal{P}_a\left[n(\xi)-\delta(\xi-x);x,t\right]-\mathcal{P}_a\left[n(\xi);x,t\right]\right\} \\
\label{eq:masterEquationDiffusiveKinaseCaLost}
\partial_t\mathcal{P}_0\left[n(\xi);t\right] &= \nu_l\int \mathrm{d}x\;\mathcal{P}_i\left[n(\xi);x,t\right],
\end{align} 
where $D_C$ and $D_K$ are, respectively, the diffusion constants of Ca$^{2+}$ and the kinase, $\nu_a$ and $\nu_d$ are the respective rates of Ca$^{2+}$ attachment to and detachment from the kinase, $\nu_l$ is the rate at which Ca$^{2+}$ is lost from the system, and $\nu_p$ again the rate at which an active kinase phosphorylates. Note, that the dynamics in the $z$-direction is irrelevant for this scenario. The initial condition is 
\begin{align}
\mathcal{P}_i\left[n(\xi)\equiv0;x,t=0\right] &= \delta(x)
\end{align}
and all other probabilities zero reflecting that the Ca$^{2+}$ enters the system at $x=0$. The probability distributions obey the normalization condition $\int\mathcal{D}n(\xi)\left\{\mathcal{P}_0+\int\mathrm{d}x\;\left[\mathcal{P}_i+\mathcal{P}_a\right]\right\}=1$ for all times $t$. In contrast to the toy model, we will scale time by $\nu_p$ and length by $\sqrt{D_C/\nu_p}$. We keep the same notation for the dimensionless parameters.

\subsubsection{Membrane-binding kinase}

The dynamics in the case of a kinase that needs to attach to a membrane for activation follows the same reasoning. The membrane is assumed to be localized at $z=0$ and to extend infinitely into the $x$-direction. Calcium enters at $x=z=0$ into the domain $z\ge0$. In addition to the distributions $\mathcal{P}_i$, when Ca$^{2+}$ is present, but not attached to the kinase, $\mathcal{P}_a$, when the kinase has Ca$^{2+}$ attached to it, but is not bound to the membrane and thus inactive, and $\mathcal{P}_0$, when Ca$^{2+}$ is lost from the system, there is the distribution $\mathcal{P}_b$, when the kinase is active, that is, it has Ca$^{2+}$ attached to it and is bound to the membrane. They obey the following Master equation:
\begin{align}
\label{eq:masterEquationMembraneKinaseInactive}
\partial_t\mathcal{P}_i\left[n(\xi);x,z,t\right] &= D_C\left(\partial_x^2+\partial_z^2\right)\mathcal{P}_i\left[n(\xi);x,z,t\right]\nonumber\\
&\quad\quad+\nu_d\mathcal{P}_a\left[n(\xi);x,z,t\right]- (\nu_a+\nu_l)\mathcal{P}_i\left[n(\xi);x,z,t\right]\\
\label{eq:masterEquationMembraneKinaseCytosolic}
\partial_t\mathcal{P}_a\left[n(\xi);x,z,t\right] &= D_K\left(\partial_x^2+\partial_z^2\right)\mathcal{P}_a\left[n(\xi);x,z,t\right]\nonumber\\
&\quad\quad -\nu_d\mathcal{P}_a\left[n(\xi);x,z,t\right] +\nu_a\mathcal{P}_i\left[n(\xi);x,z,t\right]\\
\label{eq:masterEquationMembraneKinaseActive}
\partial_t\mathcal{P}_b\left[n(\xi);x,t\right] &= \nu_b\mathcal{P}_a\left[n(\xi);x,z=0,t\right]-\nu_u\mathcal{P}_b\left[n(\xi);x,t\right]\nonumber\\
&\quad\quad +\nu_p\left\{\mathcal{P}_b\left[n(\xi)-\delta(\xi-x);x,t\right]-\mathcal{P}_b\left[n(\xi);x,t\right]\right\} \\
\label{eq:masterEquationMembraneKinaseCaLost}
\partial_t\mathcal{P}_0\left[n(\xi);t\right] &= \nu_l\int \mathrm{d}x\int_0^\infty \mathrm{d}z\;\mathcal{P}_i\left[n(\xi);x,z,t\right],
\end{align} 
where $\nu_b$ and $\nu_u$ are the respective rates of kinase binding to and unbinding from the membrane. The bulk equations are complemented by no-flux boundary conditions for the Ca$^{2+}$, that is,
\begin{align}
\left.\partial_z\mathcal{P}_i\right|_{z=0}&=0.
\end{align}
The boundary condition for the kinase accounts for its binding to and unbinding from the membrane
\begin{align}
-\left.D_K\partial_z\mathcal{P}_a\right|_{z=0}&=-\nu_b\left.\mathcal{P}_a\right|_{z=0}+\nu_u\mathcal{P}_b.
\end{align}
Again, we scale time by $\nu_p^{-1}$ and length by $\sqrt{D_C/\nu_p}$ and keep the  notation for the now dimensionless parameters.

\subsection{Number of phosphorylation events}
Let us first neglect the spatial degrees of freedom and consider only the distribution of the number of phosphorylation events. 

\subsubsection{Cytosolic kinase}
In the case, we neglect the spatial degrees of freedom, the Master equation (\ref{eq:masterEquationDiffusiveKinaseInactive})-(\ref{eq:masterEquationDiffusiveKinaseCaLost}) can be written as
\begin{align}
\label{eq:CndotCaM}
\dot{C}_n &= -\left(\nu_a+\nu_l\right)C_n + \nu_d K_n\\
\dot{K}_0 &= \nu_a C_0-\nu_d K_0-K_0\\
\label{eq:KndotCaM}
\dot{K}_n &= \nu_a C_n-\nu_d K_n-K_n+ K_{n-1}\\
\label{eq:PndotCaM}
\dot{P}_n &= \nu_l C_n,
\end{align}
where Eqs.~(\ref{eq:CndotCaM}) and (\ref{eq:PndotCaM}) hold for all $n\ge0$, whereas Eq.~(\ref{eq:KndotCaM}) is valid for $n>0$. Here, $C_n$, $K_n$, and $P_n$ denote the respective probabilities of having $n$ phosphorylation events, when the Ca$^{2+}$ ion is free, attached to the kinase, or lost from the system. 

\begin{figure}
\includegraphics[width=0.5\textwidth]{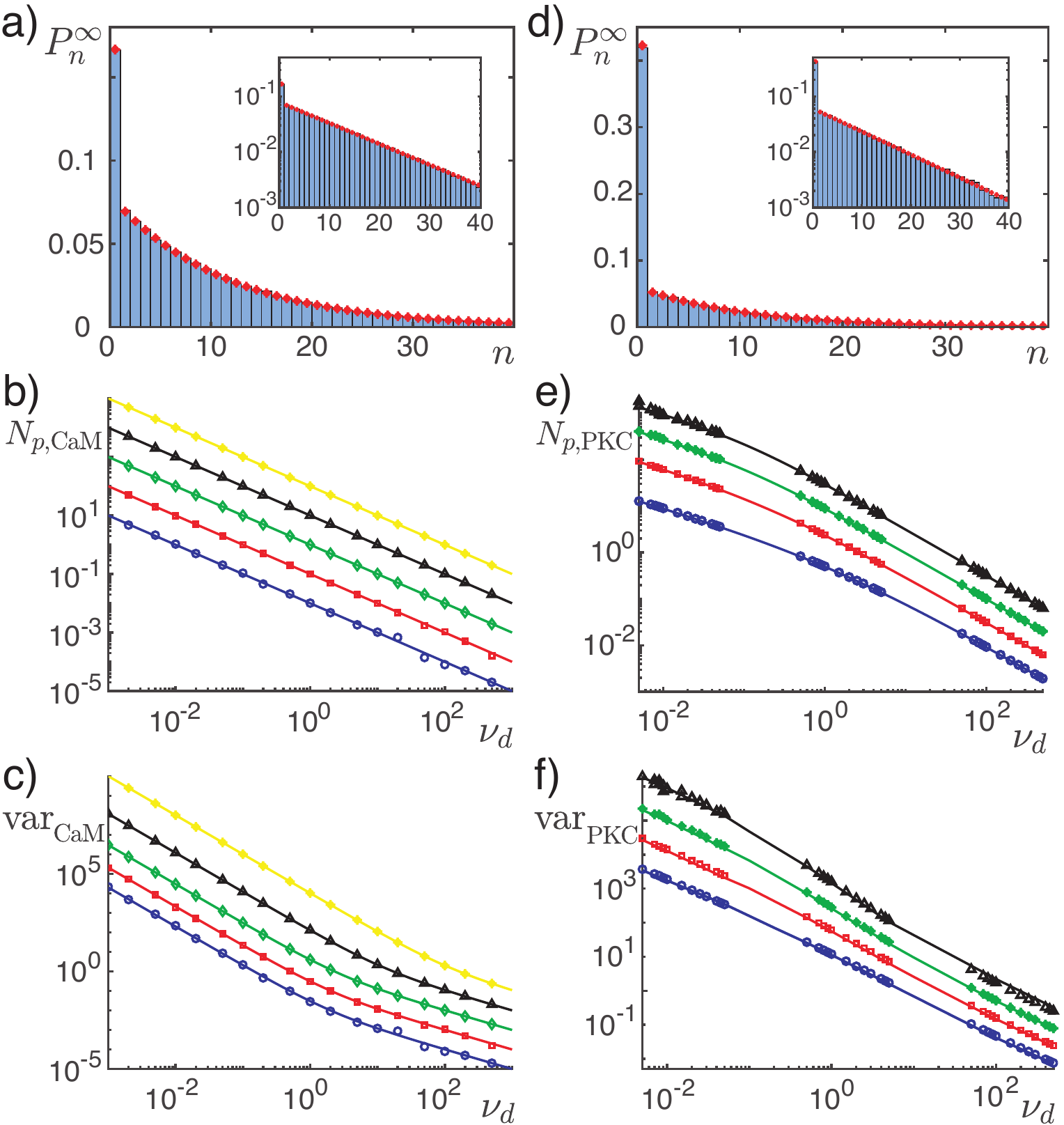}
\caption{\label{fig:phosphorylation}(color online) Number of phosphorylation events. a) The distribution $P^\infty_n$ for a cytosolic kinase with $\nu_a=\nu_d=1$ and $\nu_l=0.1$ from stochastic simulations (blue bars) and according to Eqs.~(\ref{eq:distributionPhosphorylationCaM0}) and (\ref{eq:distributionPhosphorylationCaMn}) (red stars). Inset: semilogarithmic plot of the same data. b,c) Mean number $\textcolor{black}{N_\mathrm{p,CaM}}$ (b) and variance $\textcolor{black}{\mathrm{var}_\mathrm{p,CaM}}$ (c) of phosphorylation events for a cytosolic kinase from $10^5$ simulations for each parameter set and according to Eqs.~(\ref{eq:meanPhosphorylationCaM}) and (\ref{eq:varPhosphorylationCaM}) as a function of $\nu_d$ for $\nu_a=1$ and $\nu_l=0.01$ ($\ast$, yellow), $0.1$ ($\triangle$, black), $1$ ($\diamond$, green), $10$ ($\square$, red), and $100$ ($\circ$, blue). d) The distribution $P^\infty_n$ for a membrane-binding kinase with $\nu_a=10$, $\nu_d=1$, $\nu_b=10$, $\nu_u=0.1$, $\nu_l=1$, and $D_K=1$ from stochastic simulations (blue bars) and according to Eqs.~(\ref{eq:distributionPhosphorylationPKC0}) and (\ref{eq:distributionPhosphorylationPKCn}) (red stars). Inset: semilogarithmic plot of the same data. e,f) Mean number $\textcolor{black}{N_\mathrm{p,PKC}}$ and variance $\textcolor{black}{\mathrm{var}_\mathrm{p,PKC}}$ of phosphorylation events as a function of $\nu_d$ for $\nu_a=10$, $\nu_b=\nu_u=1$, $D_K=0.01$, and $\nu_l=0.1$ ($\triangle$, black), $1$ ($\diamond$, green), $10$ ($\square$, red), and $100$ ($\circ$, blue). Lines are according to Eq.~(\ref{eq:meanPhosphorylationPKC}).}
\end{figure}

In the limit $t\to\infty$, we have $C_n=K_n=0$ for all $n\ge0$. The distribution of phosphorylation events is thus entirely determined by $P_n^\infty\equiv\lim_{t\to\infty}P_n(t)$. From Eq.~(\ref{eq:PndotCaM}) we have $P_n^\infty=\nu_l\int_0^\infty\mathrm{d}t\;C_n\equiv \nu_l\bar{C}_n$. Integrating the dynamic equations (\ref{eq:CndotCaM})-(\ref{eq:KndotCaM}) with respect to time from $t=0$ to $\infty$ and using the initial condition $C_0(t=0)=1$ and $C_{n+1}(t=0)=K_n(t=0)=0$ for all $n\ge0$ we obtain
\begin{align}
\left(\nu_a+\nu_l\right)\bar{C}_0-\nu_d \bar{K}_0 &= 1\\
\left(\nu_a+\nu_l\right)\bar{C}_n-\nu_d \bar{K}_n &= 0\\
\nu_a\bar{C}_0-\nu_d\bar{K}_0-\bar{K}_0 &= 0\\
\nu_a\bar{C}_n-\nu_d \bar{K}_n-\bar{K}_n +\bar{K}_{n-1} &=0
\end{align}
for all $n>0$. The bars indicate that the corresponding quantities have been integrated from $t=0$ to $\infty$. Solving these equations, we obtain for the distribution of the number of phosphorylation events
\begin{align}
\label{eq:distributionPhosphorylationCaM0}
P_0^\infty &= 1-\frac{\nu_a}{\nu_a+\nu_l+\nu_d\nu_l}\\
\label{eq:distributionPhosphorylationCaMn}
P_n^\infty &= \nu_a\nu_d\nu_l\frac{\left(\nu_a+\nu_l\right)^{n-1}}{\left(\nu_a+\nu_l+\nu_d\nu_l\right)^{n+1}}.
\end{align}
We present an example for the distribution of the number of phosphorylation events in Fig.~\ref{fig:phosphorylation}a.

For the average number of phosphorylation events $\textcolor{black}{N_\mathrm{p,CaM}}$ and its variance \textcolor{black}{var$_\mathrm{CaM}$} we  get
\begin{align}
\label{eq:meanPhosphorylationCaM}
\textcolor{black}{N_\mathrm{p,CaM}} &= \frac{\nu_a}{\nu_d\nu_l}\\
\label{eq:varPhosphorylationCaM}
\textcolor{black}{\mathrm{var}_\mathrm{CaM}} &= \left(1+\frac{2}{\nu_d}+\textcolor{black}{N_\mathrm{p,CaM}}\right)\textcolor{black}{N_\mathrm{p,CaM}}.%\frac{\nu_a}{\nu_d\nu_l}\left(1 + \frac{\nu_a}{\nu_d\nu_l}+\frac{2}{\nu_d}\right).
\end{align}
Compared to the toy model, the average $\textcolor{black}{N_\mathrm{p,CaM}}$ contains an extra factor $\nu_a/\nu_d$ and the variance has an additional term $2/\nu_d$. On Figure~\ref{fig:phosphorylation}b, we display the average average number of phosphorylation events as a function of the detachment rate $\nu_d$ and for several values of $\nu_l$.

\subsubsection{Membrane-binding kinase}

The calculation in the case of a membrane-binding kinase proceeds along the same lines as for the case of a cytosolic kinase. To simplify the task, we will consider the case, when neither Ca$^{2+}$ reattaches to a kinase after detaching nor the kinase rebinds after unbinding from the membrane. The Master equation can then be written as
\begin{align}
\label{eq:C0dotPKC}
\partial_t C_0 &= \partial_z^2C_0 -\left(\nu_a+\nu_l\right)C_0\\
\label{eq:K0dotPKC}
\partial_t K_0 &=D_K\partial_z^2K_0+\nu_aC_0-\nu_d K_0\\
\label{eq:k0dotPKC}
\dot k_0 &= \nu_bK_0(z=0)-\nu_uk_0-k_0\\
\label{eq:kndotPKC}
\dot k_n &= -\nu_uk_n-k_n+k_{n-1}\\
\label{eq:P0dotPKC}
\dot P_0 &=\int_0^\infty\left(\nu_lC_0+\nu_dK_0\right)\;\mathrm{d}z+ \nu_u k_0\\
\label{eq:PndotPKC}
\dot P_n &=\nu_u k_n, 
\end{align}
where Eqs.~(\ref{eq:kndotPKC}) and (\ref{eq:PndotPKC}) hold for all $n\ge1$. In these equations, $C_0$ and $K_0$, respectively, denote the probabilities of the free Ca$^{2+}$ ion and the Ca$^{2+}$ ion attached to the kinase diffusing the cytoplasm. Under the above conditions, phosphorylation cannot have occurred in these states. With $k_n$ and $P_n$ we denote the respective probabilities of having $n$ phosphorylation events with the kinase bound to the membrane and after the Ca$^{2+}$ ion is lost from the system. These equations are complemented by boundary condition for Eqs.~(\ref{eq:C0dotPKC}) and (\ref{eq:K0dotPKC}). Explicitly,
\begin{align}
\partial_z\left.C_0\right|_{z=0}&=0\\
-D_K\partial_z\left.K_0\right|_{z=0} &= -\nu_b K_0(z=0) +\nu_u k_0.
\end{align} 
Initially, the Ca$^{2+}$ ion is localized at $z=0$.

For the probability distribution $P_n^\infty$ of the number of phosphorylation events we obtain
\begin{align}
\label{eq:distributionPhosphorylationPKC0}
P_0^\infty &=1-\bar{k}_0\\
\label{eq:distributionPhosphorylationPKCn}
P_n^\infty&=\nu_u\left(1+\nu_u\right)^{-n}\bar{k}_0
\end{align}
for $n\ge1$ with
\begin{align}
\bar{k}_0 &= \frac{\nu_a\nu_b}{1+\nu_u}\frac{1}{\sqrt{\nu_a+\nu_l}}\frac{\sqrt{D_K}}{\sqrt{D_K\left(\nu_a+\nu_l\right)}+\sqrt{\nu_d}}
\frac{1}{\sqrt{D_K\nu_d}+\nu_b}.
\end{align}
Figure~\ref{fig:phosphorylation}d shows an example of the distribution.
 
From these expressions we get for the average number of phosphorylation events and the corresponding variance
\begin{align}
\textcolor{black}{N_\mathrm{p,PKC}} &= \frac{1+\nu_u}{\nu_u}\bar{k}_0\\
\textcolor{black}{\mathrm{var}_\mathrm{PKC}} &= \left(1+\frac{2}{\nu_u}-\textcolor{black}{N_\mathrm{p,PKC}}\right)\textcolor{black}{N_\mathrm{p,PKC}}.
\end{align}

Similarly, the distribution of the number of phosphorylation events can be calculated for the full Master equation (\ref{eq:masterEquationDiffusiveKinaseInactive})-(\ref{eq:masterEquationDiffusiveKinaseCaLost}), see Appendix~\ref{app:masterEquationPhosphorylationPKC}. For the mean value, we find
\begin{align}
\label{eq:meanPhosphorylationPKC}
\textcolor{black}{N_\mathrm{p,PKC}}& = \frac{\nu_b}{\nu_u} \left\{\left(\sqrt{\frac{1}{\nu_l}}+\sqrt{\frac{D_K}{\nu_d}}\right)^2%-\sqrt{\frac{D_K}{\nu_l\nu_d}}
+\frac{D_K\nu_a}{\nu_d\nu_l}\right\}^{-1/2} \frac{\nu_a}{\nu_d\nu_l}%\\
%\langle n^2\rangle-\langle n\rangle^2 &= \left(1+2\frac{\left[D_K\left(\lambda_1+\lambda_2\right)+\nu_b\right]\lambda_1\lambda_2+\nu_b\left(\nu_a+\nu_b\right)}{\nu_u\nu_dD_K\left(\lambda_1+\lambda_2\right)\lambda_1\lambda_2}-\langle n\rangle\right)\langle n\rangle.
\end{align}
The expression for the variance is very lengthy and not illuminating. The mean value and variance are shown in Figure~\ref{fig:phosphorylation}e,f as a function of $\nu_d$ and for various values of $\nu_l$, where the variance has been obtained from a numerical solution of the Master equation. %The discrepancy between the variance obtained from the numerical solution of the Master equation and the stochastic simulation is due to the finite value of $\Delta t$ used in the stochastic simulations, see Sec.~\ref{sec:stochasticSimsPKC}. In order to match the variance in the number of phosphorylation events from the stochastic simulations and the Master equation a prohibitively small value of $\Delta t$ needs to be taken. 

\subsection{Spatial distribution of phosphorylation events by a cytosolic kinase}
We now turn to the spatial distribution of phosphorylation events for a cytosolic kinase, for which we need to consider the full Master equation presented in Sect.~\ref{sec:masterEquations}. We first solve it numerically and then present results of a mean-field analysis.

\subsubsection{Stochastic simulations}
\label{sec:stochasticSimsCaM}
\begin{figure}
\includegraphics[width=0.5\textwidth]{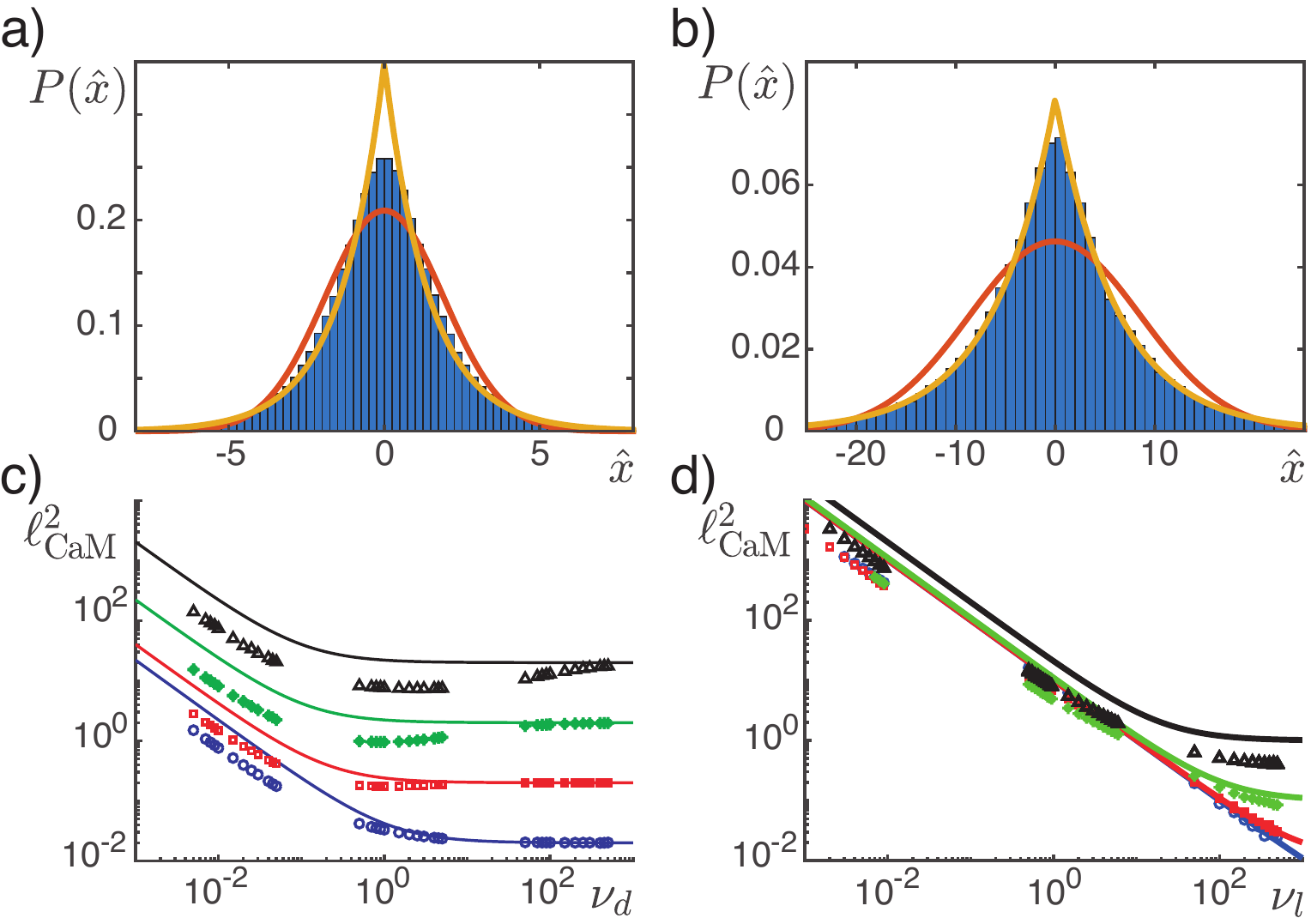}
\caption{\label{fig:CaMEstimate}(color online) Spatial distribution of phosphorylation events for a cytosolic kinase. a, b) Distributions $P(\hat{x})$ of estimated Ca$^{2+}$ entry site from $10^6$ numeric simulations for $\nu_a=1$, $\nu_d=1$ (a) and $\nu_a=10$, $\nu_d=0.1$ (b). Lines show a Gaussian (red) and an exponential ($\exp(-|x|/\lambda)/2\lambda$, yellow) fit. Other parameter values: $\nu_l=1$ and $D_K=1$. c) Estimation error as a function of $\nu_d$. d) Estimation error as a function of $\nu_l$. In (c,d) lines represent the mean-field result Eq.~(\ref{eq:errorMFCaM}). Parameter values are $\nu_a=10$, $D_K=0.01$ and $\nu_l=100$ ($\circ$, blue), $10$ ($\square$, red), $1$ ($\diamond$, green), $0.1$ ($\triangle$, black) (c) and $\nu_d=100$ ($\circ$, blue), $10$ ($\square$, red), $1$ ($\diamond$, green), $0.1$ ($\triangle$, black) (d).}
\end{figure}

The numerical analysis of the Master equation (\ref{eq:masterEquationDiffusiveKinaseInactive})-(\ref{eq:masterEquationDiffusiveKinaseCaLost}) is done through simulations as described in Sect.~\ref{sec:numericalSolution} with appropriate modifications. In Figure~\ref{fig:CaMEstimate}a, b, we present examples of the distribution $P$ of estimated positions. It is non-Gaussian and has exponential tails. 

In Figure~\ref{fig:CaMEstimate}c, we present the estimation error as a function of the detachment rate $\nu_d$. Initially, it decreases as $1/\nu_d$. For $\nu_d\approx\nu_p$, the dependence changes. For large enough values of the loss rate $\nu_l$, the error apparently saturates as a function of $\nu_d$. Below a critical loss rate, the error first increases before saturating. As a consequence, there is an optimal value of $\nu_d$ for which the error is minimal. Note, however, that this minimum is not very prominent. The dependence of the error on the loss rate $\nu_l$ is similar, see Fig.~\ref{fig:CaMEstimate}d: initially, it decreases as $1/\nu_l$ and then saturates. Saturation occurs for $\nu_l>100\nu_p$ for values of $\nu_d\lesssim 1$. With increasing values of $\nu_d$, saturation is observed for increasing values of $\nu_l$. In contrast to the dependence on $\nu_d$, our simulations do not indicate the existence of an optimum loss rate that would minimize the error.

\subsubsection{Mean-field analysis}
We now perform a mean-field analysis similar to Sect.~\ref{sec:meanfieldToyModel}. Let $p_C$ denote the probability of finding a free Ca$^{2+}$, that is, $p_C(x,t)=\int\mathcal{D}n(\xi)\;\mathcal{P}_i\left[n(\xi);x,t\right]$. Analogously, $p_K(x,t)$ denotes the probability of finding a kinase with the Ca$^{2+}$ bound at position $x$ at time $t$ and $p_0(t)$ the probability that there is no Ca$^{2+}$ in the system at time $t$. These quantities obey
\begin{align}
\partial_t p_C &= \partial_x^2 p_C + \nu_d p_K -\left(\nu_a+\nu_l\right)p_C\\
\partial_t p_K &= D_K\partial_x^2 p_K -\nu_d p_K + \nu_a p_C\\
\dot{p}_0 &= \nu_l\int \mathrm{d}x\; p_C.
\end{align}
The normalization condition reads $\int\left(p_C+p_K\right)\;\mathrm{d}x + p_0=1$ and the initial condition is $p_C(x,t=0)=\delta(x)$. 

Let us focus on the limit $t\to\infty$. Similar to the mean-field analysis of the toy model, the distribution $P$ of having $n$ phosphorylation events at $x$ is given by 
\begin{align}
P(n,x)&=\frac{1}{n!}\bar{p}_K(x)^n\exp\left\{-\bar{p}_K(x)\right\},
\end{align}
where the barred quantities indicate as above time-integrated quantities, for example, $\bar{p}_{K}(x)=\int_0^\infty p_{K}(x,t)\;\mathrm{d}t$. To obtain the distribution of expected phosphorylation events, we are then left with solving
\begin{align}
-1 &= \partial_x^2 \bar{p}_C +\nu_d\bar{p}_K-\left(\nu_a+\nu_l\right)\bar{p}_C\\
0 &= D_K\partial_x^2 \bar{p}_K -\nu_d\bar{p}_K+\nu_a\bar{p}_C.
\end{align}  
In Fourier space, we obtain
\begin{align}
\bar{p}_{K,q} &= \nu_a\left\{D_Kq^4+\left[D_K\left(\nu_a+\nu_l\right)+\nu_d\right]q^2+\nu_l\nu_d\right\}^{-1}.
\end{align}
We do not need the expression for $\bar{p}_{C,q}$, as the expected $\hat{n}$ number of phosphorylation events in the limit $t\to\infty$ is $\hat{n}(x)=\bar{p}_K(x)$. Using $\int\hat{n}(x)\;dx=\bar{p}_{K,0}$ and $\int x^2\hat{n}(x)\;dx=-\bar{p}^{\prime\prime}_{K,0}$ as well as expression (\ref{eq:errorMF}) for determining the estimation error, we finally obtain for the error in the mean-field limit
\begin{align}
\label{eq:errorMFCaM}
\ell_\mathrm{CaM}^2 = 2\left\{\ell_C^2 +  \ell_K^2\left(1+\frac{\nu_a}{\nu_l}\right)\right\},
\end{align}
where $\ell_C^2=1/\nu_l$ is the diffusion length of Ca$^{2+}$ and $\ell_K^2=D_K/\nu_d$ (we recall that in the rescaled units used here, $\nu_p=D_C=1$). 

The mean-field result reproduces some of the features presented by the simulation results, see Fig.~\ref{fig:CaMEstimate}c,d: As for the stochastic simulations, the error decays inversely proportional with the loss- and the detachment rates if $\nu_l,\nu_d\ll1$. Furthermore, the error saturates if these rates are large with $\ell_\mathrm{CaM}^2\to2 D_K/\nu_d$ for $\nu_l\to\infty$ and $\ell_\mathrm{CaM}^2\to2/\nu_l$ for $\nu_d\to\infty$. However, neither as a function of $\nu_d$ nor of $\nu_l$ does the mean-field calculation indicate optimal rates that would minimize the error. \textcolor{black}{} The mean-field estimation error agrees quantitatively with the simulation result in the limits of small loss rates $\nu_l$ and large detachment rates $\nu_d$. 

\subsection{Spatial distribution of phosphorylation events by a membrane-binding kinase}

\subsubsection{Stochastic simulations}
\label{sec:stochasticSimsPKC}
In the case of a membrane-binding kinase, the presence of a boundary at $z=0$ requires special attention in the stochastic simulation. Whenever the system is in a state, where the Ca$^{2+}$ ion is diffusing in the cytoplasm, or when the kinase is bound to the membrane and active, we use a Gillespie-like algorithm as explained for the toy model. If the kinase is bound to the membrane at $x$ and thus active, it can either phosphorylate or unbind from the membrane. In the first case, we record the position of the phosphorylation event, otherwise the system state is changed and the new coordinates of the now unbound kinase are $(x,0)$. If the system is in a state of an unattached Ca$^{2+}$ ion, then the new position $(x_\mathrm{new},z_\mathrm{new})$ is determined as in Sec.~\ref{sec:numericalSolution}. Should $z_\mathrm{new}<0$, which is outside the considered domain, then the particle is assumed to have been reflected and the $z$ coordinate of the Ca$^{2+}$ is set to $-z_\mathrm{new}$. 

In case, the Ca$^{2+}$ ion is attached to the kinase, which itself is residing in the cytoplasm, then binding to the membrane needs to be considered. To do so, we use in this state a scheme with continuous space and discrete time steps of length $\Delta t$ and employ the methods presented in Ref.~\cite{Erban:2007fn}: During each time step, we first determine the new position $(x_\mathrm{new},z_\mathrm{new})$ of the particle at $t+\Delta t$, as explained in Sec.~\ref{sec:numericalSolution}. If $z_\mathrm{new}<0$, then the kinase has ``crossed'' the membrane and one has to determine, whether it bound to the membrane during this process. To this end, a new random number between 0 and 1 is drawn. If it is smaller than $1-\nu_b\sqrt{\pi \Delta t}/(2\sqrt{D_K})$, then the kinase has not bound to the membrane, but instead was reflected and the new position is $(x_\mathrm{new},-z_\mathrm{new})$. We then determine if the Ca$^{2+}$ has detached and change the state if necessary. In the opposite case, the kinase binds to the membrane at $(x_\mathrm{new},0)$ and the system state is changed accordingly. Even if $z_\mathrm{new}>0$, the kinase might still have bound to the membrane. To determine, whether this happened a random number between 0 and 1 is drawn. If it is smaller than $\exp\left\{-(z_\mathrm{new}z_\mathrm{old})/(D_K*\Delta t)\right\}\nu_b\sqrt{\pi \Delta t}/(2\sqrt{D_K})$, then the kinase bound to the membrane at $(x_\mathrm{new},0)$~\cite{Andrews:2004fs}. If the kinase has not bound to the membrane, we check whether the Ca$^{2+}$ detached and change the state if necessary. The size of the time step $dt$ is chosen to be $\Delta t=0.1/\max\{\nu_b,\nu_d\}$.

\begin{figure}
\includegraphics[width=0.5\textwidth]{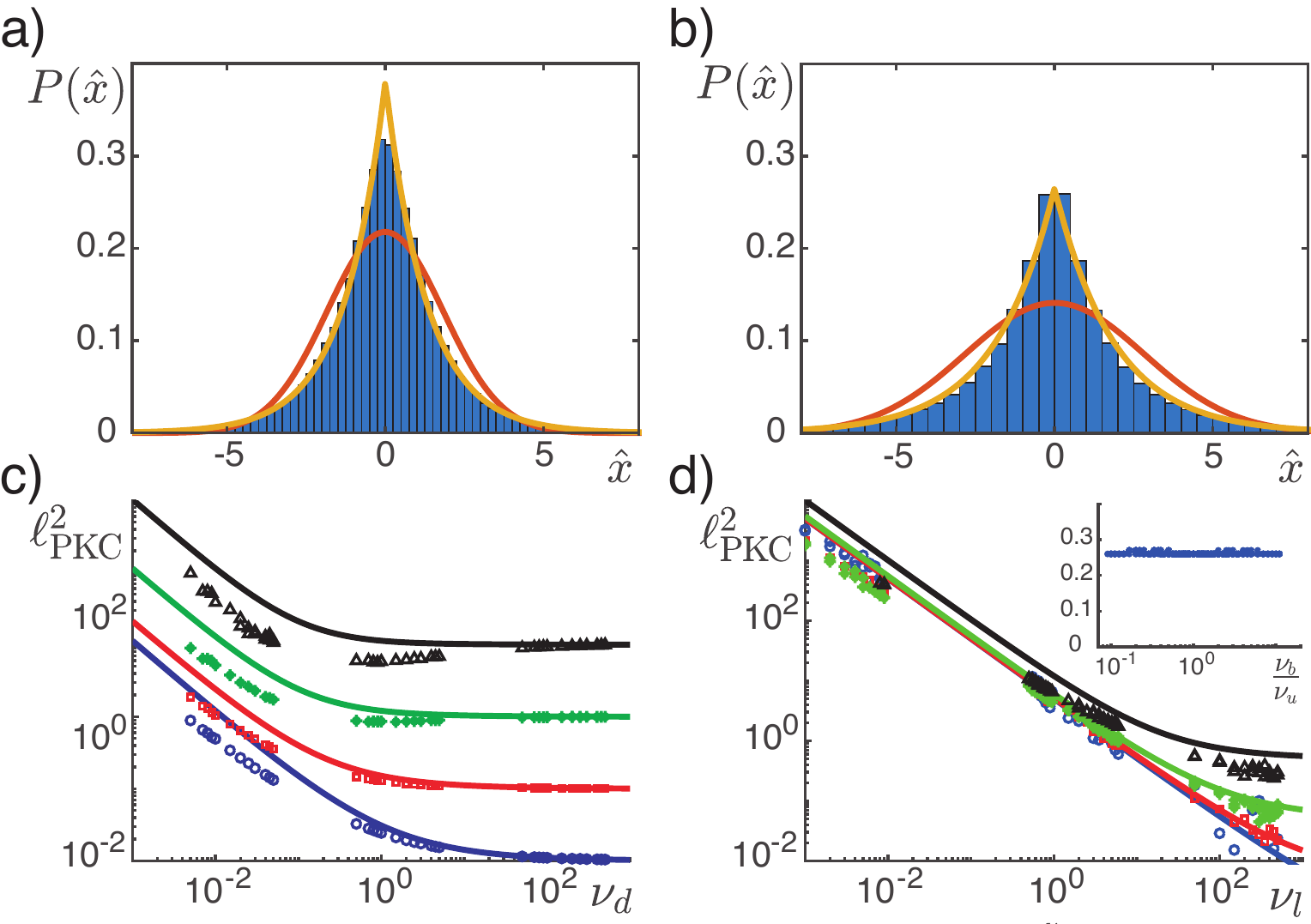}
\caption{\label{fig:PKCEstimate}(color online) Spatial distribution of phosphorylation events for a membrane-binding kinase. a, b) Distributions $P(\hat{x})$ of estimated Ca$^{2+}$ entry site from $10^6$ numeric simulations for $\nu_a=1$ (a) and $\nu_a=10$ (b). Lines show a Gaussian (red) and an exponential ($\exp(-|x|/\lambda)/2\lambda$, yellow) fit. Other parameter values: $\nu_d=1$, $\nu_l=1$, $\nu_b=1$, $\nu_u=1$, and $D_K=1$. c) Estimation error as a function of $\nu_d$. d) Estimation error as a function of $\nu_l$. Inset: estimation error as a function of $\nu_b/\nu_u$. In (c,d) lines represent the mean-field result Eq.~(\ref{eq:errorMFCaM}). Parameter values are $\nu_a=10$, $D_K=0.01$ and $\nu_l=100$ ($\circ$, blue), $10$ ($\square$, red), $1$ ($\diamond$, green), $0.1$ ($\triangle$, black) (c) and $\nu_d=100$ ($\circ$, blue), $10$ ($\square$, red), $1$ ($\diamond$, green), $0.1$ ($\triangle$, black) (d).}
\end{figure}

In Figure~\ref{fig:PKCEstimate}a, b two examples of the distribution of estimated positions are shown. As in the previous cases, the distributions are not Gaussian, but instead have an exponential tail. The dependence of the estimation error on the detachment rate $\nu_d$ and the loss rate $\nu_l$ are shown in Fig.~\ref{fig:PKCEstimate}c,d. Overall, the behavior is similar to the case of a cytosolic kinase: After an initial decrease of the error with $\nu_d$ and $\nu_l$, the error saturates. As a function of $\nu_d$, saturation occurs around $\nu_d\approx\nu_p$. In contrast to the cytosolic kinase, a clear minimum of the error as a function of $\nu_d$ cannot be detected even for small loss rates. Finally, let us note that the estimation error is independent of the membrane binding and unbinding rates $\nu_b$ and $\nu_u$, as long as they have finite values, see Fig.~\ref{fig:PKCEstimate}d, inset.

\subsubsection{Mean-field analysis}
The mean-field analysis proceeds along the same lines as for the cytosolic kinase. As above, let $p_C$ denote the probability of finding a free Ca$^{2+}$, $p_C(x,z,t)=\int\mathcal{D}n(\xi)\;\mathcal{P}_i\left[n(\xi);x,z,t\right]$. Analogously, $p_K$ denotes the probability distribution for a Ca$^{2+}$ bound kinase in the cytosol, $p_k$ the one for the membrane-bound kinase, and $p_0(t)$ the probability that the Ca$^{2+}$ is lost from the system at time $t$. These quantities obey
\begin{align}
\label{eq:dpCdtPKC}
\partial_t p_C &= \left(\partial_x^2+\partial_z^2\right) p_C + \nu_d p_K -\left(\nu_a+\nu_l\right)p_C\\
\partial_t p_K &= D_K\left(\partial_x^2+\partial_z^2\right) p_K -\nu_d p_K + \nu_a p_C\\
\label{eq:dpkdtPKC}
\partial_t p_k &= \nu_b p_K(z=0) - \nu_u p_k\\
\label{eq:dp0dtPKC}
\dot{p}_0 &= \nu_l\int \mathrm{d}x\int \mathrm{d}z\; p_C.
\end{align}
These equations are complemented by boundary conditions at the membrane. Explicitly, $\partial_z\left.p_C\right|_{z=0}=0$ and $-D_K\partial_z\left.p_K\right|_{z=0}=-\nu_bp_K(z=0)+\nu_up_k$. The normalization condition reads $\int\mathrm{d}x\int\mathrm{d}z\left(p_C+p_K\right)+\int p_k \;\mathrm{d}x+ p_0=1$ and the initial condition is $p_C(x,z,t=0)=\delta(x)\delta(z)$. Finally, the distribution of phosphorylation events in space for $t\to\infty$ is given by $\hat{n}(x)=\int_0^\infty p_k(x,t)\;\mathrm{d}t$.

To obtain the latter, we first integrate Eqs.~(\ref{eq:dpCdtPKC})-(\ref{eq:dpkdtPKC}) with respect to time, which yields
\begin{align}
-\delta(x)\delta(z) &= \left(\partial_x^2+\partial_z^2\right) \bar{p}_C + \nu_d \bar{p}_K -\left(\nu_a+\nu_l\right)\bar{p}_C\\
0 &= D_K\left(\partial_x^2+\partial_z^2\right) \bar{p}_K -\nu_d \bar{p}_K + \nu_a \bar{p}_C\\
\label{eq:barpkPKC}
0 &= \nu_b \bar{p}_K(z=0) - \nu_u \bar{p}_k,
\end{align}
where the bars indicate the time-integrated quantities as above. From Eq.~(\ref{eq:barpkPKC}) the boundary condition for $\bar{p}_K$ at $z=0$ is seen to be $\partial_z\left.\bar{p}_K\right|_{z=0}=0$. Furthermore, it shows that $\hat{n}(x)=\nu_b \bar{p}_K(x,z=0)/\nu_u$. The solution for $\bar{p}_C$ and $\bar{p}_K$ is easiest obtained after performing a Fourier transform with respect to $x$ and a cosine transform with respect to $z$. It yields
\begin{align}
-1 &= -\left(q^2+k^2\right)\bar{p}_{C,qk}+\nu_d \bar{p}_{K,qk}-\left(\nu_a+\nu_l\right)\bar{p}_{C,qk}\\
0 & = -D_K\left(q^2+k^2\right)\bar{p}_{K,qk}-\nu_d \bar{p}_{K,qk}+\nu_a\bar{p}_{C,qk},
\end{align}
where the indices $q$ and $k$ denote the wavenumbers in $x$- and $z$-direction, respectively. The solution for the time-integrated distribution of the cytosolic kinase bound to Ca$^{2+}$ is
\begin{align}
\bar{p}_{K,qk} &= \nu_a\left\{\left[D_K\left(q^2+k^2\right)+\nu_d\right]\left(q^2+k^2+\nu_a+\nu_l\right)-\nu_d\nu_a\right\}^{-1}.
\end{align}
From this expression, we eventually get for the error
\begin{align}
\label{eq:errorMFPKC}
\ell_\mathrm{PKC}^2 &= \frac{1}{2}\left[\ell_\mathrm{CaM}^2 + \ell_C\ell_K\right].
\end{align}
Using this expression for the estimation error, we can write the average number of phosphorylation events, Eq.~(\ref{eq:meanPhosphorylationPKC}) as
\begin{align}
\textcolor{black}{N_\mathrm{p,PKC}}\equiv\int\hat{n}(x)\;\mathrm{d}x & = \frac{\nu_b}{\nu_u} \left[2\ell^2_\text{PKC}+\ell_C\ell_K\right]^{-1/2}\frac{\nu_a}{\nu_d\nu_l}.
\end{align}

\subsection{Comparison between the two scenarios}

\begin{figure}
\includegraphics[width=0.5\textwidth]{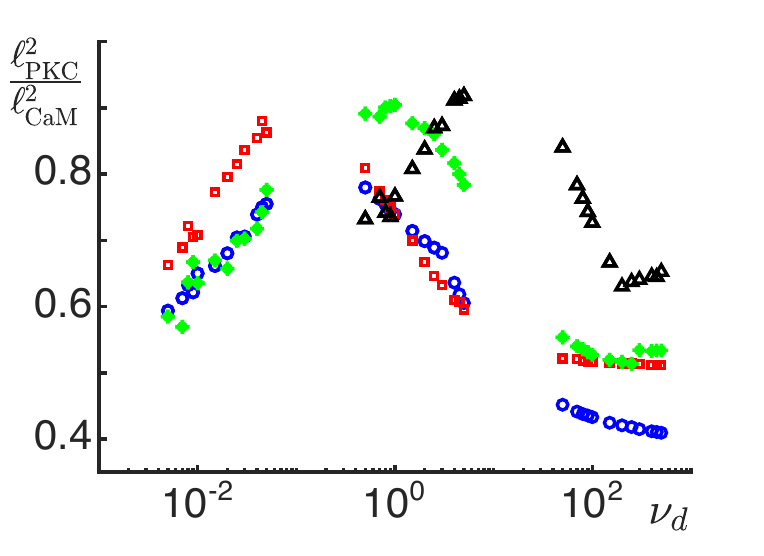}
\caption{\label{fig:ratioErrorEstimates}(color online) Ratio of the estimation error $\ell_\mathrm{CaM}^2$ of a cytosolic and $\ell_\mathrm{PKC}^2$ of a membrane-binding kinase as a function of $\nu_d$. Parameter values are $\nu_a=10$, $\nu_b=1$, $\nu_u=1$, $D_K=0.01$ and \textcolor{black}{$\nu_l=0.1$ ($\triangle$, black), $1$ ($\diamond$, green), $10$ ($\square$, red), $100$ ($\circ$, blue)}.}
\end{figure}

The simulation results show that the estimation error of the measured position for the cytosolic kinase is always larger than for the membrane-binding kinase, if we compare simulations with the same parameter values, see Fig.~\ref{fig:ratioErrorEstimates}. This result also obtained by the mean-field expressions for $\ell_\mathrm{CaM}$ and $\ell_\mathrm{PKC}$. However, the mean-field result for the estimation error ratio does not represent the functional dependence of the ratio on $\nu_d$ well. Instead, it is rather constant with a value that is close to the maximal value of the ratio in the interval of $\nu_d$ displayed on Fig.~\ref{fig:ratioErrorEstimates}. 

Since for a membrane-binding kinase the estimation error is independent of the membrane binding and unbinding rates, $\nu_b$ and $\nu_u$, we cannot meaningfully compare the number of phosphorylation events for the two scenarios. Note, however, that by increasing the ratio $\nu_b/\nu_u$, the number of phosphorylation events can be increased without affecting the accuracy of position estimate in the case of a membrane-binding kinase. This is another advantage of membrane-binding kinases over cytosolic kinases: signals will be transmitted with higher fidelity in case more target proteins are phosphorylated. %For cytosolic kinase, the general trend indicates that the estimation error decreases with decreasing number of phosphorylation events.

%\textcolor{black}{Finally,  the results for the probabilities $P_0^\infty$ of having no phosphorylation event, Eqs.~\eqref{eq:distributionPhosphorylationCaM0} and \eqref{eq:distributionPhosphorylationPKC0}, show that, for the same values of the rates $\nu_a$, $\nu_d$, and $\nu_l$ and the diffusion constant $D_K$, and arbitrary values of the rates $\nu_b$ and $\nu_u$, a membrane-binding kinase is always more likely to not lead to any phosphorylation event than a cytosolic kinase, $P_{0,\mathrm{PKC}}^\infty<P_{0,\mathrm{CaM}}^\infty$, where the indices 'CaM' and 'PKC', respectively, refer to the CaM and the PKC scenarios. These probabilities both increase monotonically with $\nu_d$ and $\nu_l$ and tend to one in the limits $\nu_l\to\infty$ and $\nu_d\to\infty$.}

\section{Responses to Ca$^{2+}$ puffs}
\label{sec:puffs}
We will now turn to situations, in which more than one Ca$^{2+}$ is present in the system. Since all particles are independent of each other, one might expect that the error $\ell^2$ simply scales as $N_\mathrm{Ca}^{-1/2}$ if $N_\mathrm{Ca}$ is the number of Ca$^{2+}$ ions. However, since only those events are counted in which a phosphorylation took place, this expectation is not met. Following the presentation of simulation results, we will apply a mean-field \textit{ansatz} to express the error for a puff in terms of the average phosphorylation profile $\hat{n}$ for a single Ca$^{2+}$ ion.

\subsection{Stochastic simulations}

Simulations are done as described in Sects.~\ref{sec:stochasticSimsCaM} and \ref{sec:stochasticSimsPKC}. For a puff of $N_\mathrm{Ca}$ Ca$^{2+}$ ions, we ran $N$ simulations and recorded the positions of all phosphorylation events during these simulations. We then obtained the estimated position by calculating their average. For each data point we performed at least $2\cdot10^6$ simulations.

\begin{figure}
\includegraphics[width=0.5\textwidth]{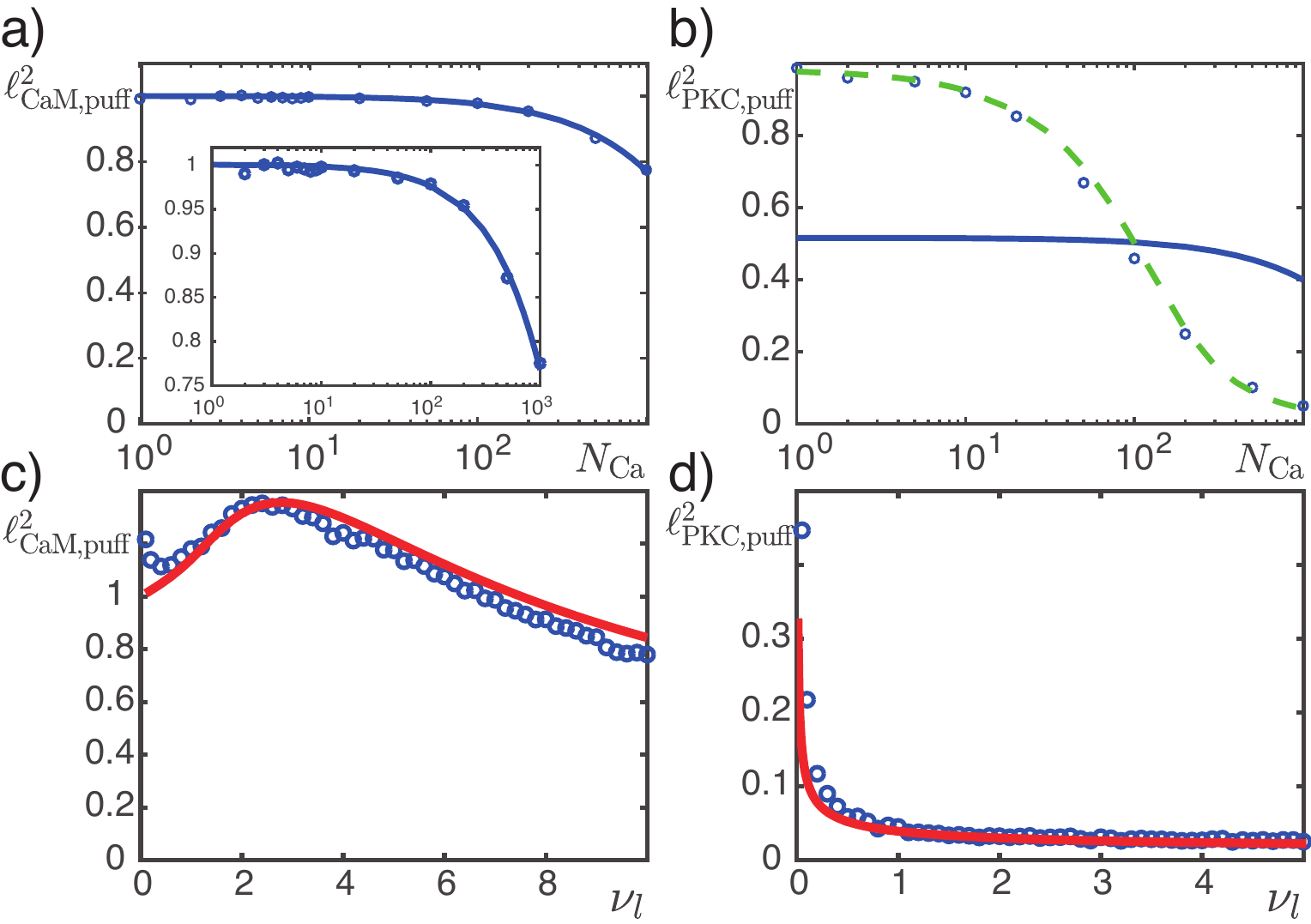}
\caption{\label{fig:puffNCa}(color online) Estimation error for a Ca$^{2+}$ ion puff. a, b) Estimation error as a function of the number $N_\mathrm{Ca}$ of Ca$^{2+}$ ions in a puff from simulations (blue circles) and the mean-field result (\ref{eq:estimationErrorMF}) (blue line) for a cytosolic (a) and a membrane-binding kinase (b). The green dashed line in (b) is a fit of Eq.~(\ref{eq:estimationErrorMF}) to the simulation data with rescaled parameters $\ell^2$ and $\langle n\rangle_1$. c, d) Estimation error for a puff of $N_\mathrm{Ca}=1000$ Ca$^{2+}$ ions as a function of the loss rate from simulations (blue circles) and Eq.~(\ref{eq:estimationErrorMF}) (red full line) for a cytosolic (c) and a membrane-binding kinase (d). Parameters are $D_K=0.01$, $\nu_a=1$ (a,d), $\nu_a=10$ (b), and $\nu_a=0.01$ (c), $\nu_d=100$ (a) and $\nu_d=1$ (b-d), $\nu_l=10$ (a,b), $\nu_b=1$ (b,d), and $\nu_u=1$ (b,d).}
\end{figure}

In Figure~\ref{fig:puffNCa}a,b, we present the estimation error for a puff as a function of the number $N_\mathrm{Ca}$ of Ca$^{2+}$ ions per puff. For both kinds of kinases, the estimation error decreases monotonically with increasing $N_\mathrm{Ca}$. The data points fall onto a sigmoidal curve: initially the accuracy of the estimate increases less than for larger values of $N_\mathrm{Ca}$. Note, that with increasing number of Ca$^{2+}$ ions, the gap between the estimation error for the membrane-binding kinase and the cytosolic kinase gets wider. 

In the case of a cytosolic kinase and for a Calcium puff with $N_\mathrm{Ca}=1000$, the dependence of the estimation error as a function of $\nu_l$ changes qualitatively with respect to the case of a single Ca$^{2+}$ ion: In the latter case the error decreases monotonically whereas for the puff a minimum \textcolor{black}{and a maximum} exist, \textcolor{black}{respectively,} around $\nu_l=1$ \textcolor{black}{and $2$}, see Fig.~\ref{fig:puffNCa}c. In contrast, for a membrane-binding kinase the dependence remains monotonically decreasing, see Fig.~\ref{fig:puffNCa}d. However, the estimation error saturates for smaller values of $\nu_l$ compared to the case of a single Ca$^{2+}$ ion.

\subsection{Mean-field analysis}
\label{sec:puffMF}
We will now apply the mean-field \textit{ansatz} introduced in Sect.~\ref{sec:meanfieldToyModel} to the case of Ca$^{2+}$ puffs, where we focus directly on the limit $t\to\infty$. For the general expression~(\ref{eq:estimationError2}) for the estimation error, we need the probability $\mathcal{P}\left[n(\xi)\right]$ for a particular realization of the phosphorylation profile $n(\xi)$. In the mean-field approximation, the probability $P(n,\xi)$ of having $n$ phosphorylation events at position $\xi$ is given by Eq.~(\ref{eq:probabilityMF}), where in the general case $\bar{p}_a$ has to be replaced by the time-integrated probability to find the particle in the phosphorylating state at position $\xi$. Explicitly, for a cytosolic kinase it is $\bar{p}_K$, whereas for the membrane-binding kinase it is $\bar{p}_k$. In contrast to the case of a single Ca$^{2+}$ ion, we cannot use expression~(\ref{eq:errorMF}) for the error, because it does not depend on the average number of phosphorylation events $\langle n\rangle$, which obviously increases with the number of Ca$^{2+}$ ions $N_\mathrm{Ca}$. Therefore, we will now calculate the exact expression of the estimation error~(\ref{eq:estimationError2}) in the mean-field approximation.  

From $P(n,\xi)$, we obtain the mean-field probability $\mathcal{P}_\mathrm{mf}\left[n(\xi)\right]$ of a particular realization $n(\xi)$ through
\begin{align}
\mathcal{P}_\mathrm{mf}\left[n(\xi)\right] &= \mathcal{N}\prod_\xi P(n,\xi)
\end{align}
for all $n(\xi)$ with $\int\mathrm{d}\xi\;n(\xi)\neq0$. The normalization factor $\mathcal{N}$ assures that $\int \mathcal{D}' n(\xi) \mathcal{P}\left[n(\xi)\right]=1$, where the prime indicates summation over all distributions $n(\xi)$ except for $n(\xi)\equiv0$. Explicitly, 
\begin{align}
\mathcal{P}_\mathrm{mf}\left[n(\xi)\right] 
&= \frac{\mathrm{e}^{-\int \mathrm{d}\xi'\;\hat{n}(\xi')}}{1-\mathrm{e}^{-\int \mathrm{d}\xi'\;\hat{n}(\xi')}}\prod_x\frac{1}{n(\xi)!}\left(\hat{n}(\xi)\right)^{n(\xi)},\\
\intertext{where $\hat{n}(\xi)$ is the average phosphorylation profile, such that}
\mathcal{P}_\mathrm{mf}\left[n(\xi)\right]&= \frac{\mathrm{e}^{-N_\mathrm{p}}}{1-\mathrm{e}^{-N_\mathrm{p}}}\prod_\xi\frac{1}{n(\xi)!}\left(\hat{n}(\xi)\right)^{n(\xi)}.
\end{align}
Using this probability distribution in expression~(\ref{eq:estimationError2}) for the estimation error, we get
\begin{align}
\ell_\mathrm{puff}^2&= \int\mathcal{D}'n(\xi)\frac{\int \mathrm{d}\xi_1\int\mathrm{d}\xi_2\;\xi_1\xi_2n(\xi_1)n(\xi_2)}{\left(\int\mathrm{d}\xi'\;n(\xi')\right)^2}\mathcal{P}\left[n(\xi)\right]\\
&= \frac{\mathrm{e}^{-N_\mathrm{p}}}{1-\mathrm{e}^{-N_\mathrm{p}}}\prod_\xi\sum_{n(\xi)=0}^\infty\frac{\int \mathrm{d}\xi_1\int\mathrm{d}\xi_2\;\xi_1\xi_2n(\xi_1)n(\xi_2)}{\left(\int\mathrm{d}\xi'\;n(\xi')\right)^2}\frac{1}{n(\xi)!}\hat{n}(\xi)^{n(\xi)}\\
&= \frac{\mathrm{e}^{-N_\mathrm{p}}}{1-\mathrm{e}^{-N_\mathrm{p}}}\prod_\xi\sum_{n(\xi)=0}^\infty\int\mathrm{d}\xi_1\int\mathrm{d}\xi_2\;\xi_1\xi_2\hat{n}(\xi_1)\frac{\delta}{\delta\hat{n}(\xi_1)}\hat{n}(\xi_2)\frac{\delta}{\delta\hat{n}(\xi_2)} \frac{\hat{n}(\xi)^{n(\xi)}}{n(\xi)!}\frac{1}{\left(\int\mathrm{d}\xi'\;n(\xi')\right)^2}\\
&= \frac{\mathrm{e}^{-N_\mathrm{p}}}{1-\mathrm{e}^{-N_\mathrm{p}}}\int\mathrm{d}\xi_1\int\mathrm{d}\xi_2\;\xi_1\xi_2\hat{n}(\xi_1)\frac{\delta}{\delta\hat{n}(\xi_1)}\hat{n}(\xi_2)\frac{\delta}{\delta\hat{n}(\xi_2)}\prod_\xi\sum_{n(\xi)=0}^\infty \frac{\hat{n}(\xi)^{n(\xi)}}{n(\xi)!}\frac{1}{\left(\int\mathrm{d}\xi'\;n(\xi')\right)^2}\\
&= \frac{\mathrm{e}^{-N_\mathrm{p}}}{1-\mathrm{e}^{-N_\mathrm{p}}}\int\mathrm{d}\xi_1\;\xi_1^2\hat{n}(\xi_1)\frac{\delta}{\delta\hat{n}(\xi_1)} \prod_\xi\sum_{n(\xi)=0}^\infty\frac{\hat{n}(\xi)^{n(\xi)}}{n(\xi)!}\frac{1}{\left(\int\mathrm{d}\xi'\;n(\xi')\right)^2}\\
&= \frac{\mathrm{e}^{-N_\mathrm{p}}}{1-\mathrm{e}^{-N_\mathrm{p}}}\int\mathrm{d}\xi_1\;\xi_1^2\hat{n}(\xi_1)\frac{\delta}{\delta\hat{n}(\xi_1)} \sum_{N=1}^\infty\frac{\left(\int\mathrm{d}\xi\;\hat{n}(\xi)\right)^N}{N!N^2}\\
&=\frac{\mathrm{e}^{-N_\mathrm{p}}}{1-\mathrm{e}^{-N_\mathrm{p}}}\frac{\int\mathrm{d}\xi\;\xi^2\hat{n}(\xi)}{\int\mathrm{d}\xi\;\hat{n}(\xi)}\sum_{N=1}^\infty\frac{\left(\int\mathrm{d}\xi\;\hat{n}(\xi)\right)^N}{N!N}\\
\label{eq:estimationErrorMF}
&=\ell^2\frac{\mathrm{e}^{-N_\mathrm{p}}}{1-\mathrm{e}^{-N_\mathrm{p}}}\sum_{N=1}^\infty\frac{N_\mathrm{p}^N}{N!N}.
\end{align}
In the final expression $\ell^2$ is the estimation error in the mean-field approximation, see Eq.~(\ref{eq:errorMF}). For $N_\mathrm{p}\ll1$, we have $\ell_\mathrm{puff}^2=\ell^2$ as announced in Sect.~\ref{sec:meanfieldToyModel}. In the case of a Ca$^{2+}$ puff with $N_\mathrm{Ca}$ Ca$^{2+}$ ions, $N_\mathrm{p}=N_\mathrm{Ca}\langle n\rangle_1$, where $\langle n\rangle_1$ is the average number of phosphorylation events for one Ca$^{2+}$ ion, because we assume that all Ca$^{2+}$ ions are independent of each other and that there is an excess of kinases. As the final expression for $\ell_\mathrm{puff}^2$ shows, the dependence on $N_\mathrm{Ca}$ is more complicated than the usual $1/N_\mathrm{Ca}$-dependence of the variance in case one has $N_\mathrm{Ca}$ independent measurements. The reason is that not all Ca$^{2+}$ ions produce at least one phosphorylation event, such that the position of Ca$^{2+}$ influx cannot be estimated for all Ca$^{2+}$ ions. However, for $N_\mathrm{Ca}\gg1$, we find again $\ell_\mathrm{puff}^2\sim\frac{\ell^2}{N_\mathrm{Ca}}$.

We will now use this general expression in combination with the results for $\hat{n}(x)$ obtained from the mean-field analysis. For the case of a cytosolic kinase, we obtain
\begin{align}
\label{eq:errorCaMpuff}
\ell_\mathrm{CaM,puff}^2 &= 2\left\{\ell_C^2 +  \ell_K^2\left(1+\frac{\nu_a}{\nu_l}\right)\right\}\frac{\mathrm{e}^{-\frac{N_\mathrm{Ca}\nu_a}{\nu_d\nu_l}}}{1-\mathrm{e}^{-\frac{N_\mathrm{Ca}\nu_a}{\nu_d\nu_l}}}\sum_N\frac{1}{N!N}\frac{N_\mathrm{Ca}^N\nu_a^N}{\nu_d^N\nu_l^N}.
\end{align}
Application of the general expression (\ref{eq:estimationErrorMF}) to the case of a membrane-binding kinase is tricky. This expression depends on $N_\mathrm{p,PKC}$, which in turn can be changed by changing either $N_\mathrm{Ca}$ or $\nu_b/\nu_u$. In the simulations, however, only changing $N_\mathrm{Ca}$ affects the estimation error, whereas it is independent of the ration $\nu_b/\nu_u$ as we had seen above for the case of a single Ca$^{2+}$ ion. Since these two effects are not separated in the mean-field expression one cannot expect it to describe the dependence of the error on the number of Ca$^{2+}$ ions. We thus refrain from giving the mean-field result for the case of a membrane-binding kinase. 

In Figure~\ref{fig:puffNCa}, we present the estimation error for a puff obtained from the mean-field treatment as a function of $N_\mathrm{Ca}$. In case of a cytosolic kinase, where the error for a single Ca$^{2+}$ ion is given by the mean-field result, the dependence on the number of Ca$^{2+}$ matches the simulation results perfectly. From Equation~(\ref{eq:errorCaMpuff}), we see that the error only decreases significantly, when $N_\mathrm{Ca}\sim \nu_d\nu_l/\nu_a$. For the parameters chosen in Fig.~\ref{fig:puffNCa}, we get $\nu_d\nu_l/\nu_a=1000$, which matches well the simulation data. As we have argued before, we cannot expect the mean-field error for puffs to describe simulation results for a membrane-binding kinase. For the parameters chosen in Fig.~\ref{fig:puffNCa}b, it is indeed off. However, by appropriately rescaling $\ell^2$ for $N_\mathrm{Ca}=1$ and $\langle n\rangle_1$, the expression (\ref{eq:estimationErrorMF}) provides a fit to the data. In Figure~\ref{fig:puffNCa}d, we see that for increasing values of $\nu_l$, we obtain agreement between Eq.~(\ref{eq:estimationErrorMF}) and the simulation results. These results show that the mean-field expression does capture important aspects of the estimation error even in the case of a membrane-binding kinase.

\section{Estimating the site of Ca$^{2+}$ release in presence of background phosphorylation}
Living cells have a cytosolic Ca$^{2+}$ concentration of roughly 100 nM~\cite{Milo:2015uq}. Consequently, a fraction of calmodulin and PKC$\alpha$ are active even in absence of an external signal. How does the corresponding background phosphorylation affect the accuracy of the estimated position of the Ca$^{2+}$ release site? On general grounds, cells might be expected to suppress the influence of the background by employing a threshold mechanism: a cellular response is only elicited if the number of phosphorylation events exceeds a certain value. Still, it is interesting to account explicitly for background phosphorylation in our analysis. 

In the presence of background phosphorylation, our theoretic approach has to be modified to some extent. Above, we considered the distribution of all phosphorylation sites that were generated by a Ca$^{2+}$ ion or -puff, independently of when they occurred. If we applied the same approach in presence of background phosphorylation, then the background would always outcompete the signal. We thus introduce a rate $\nu_\mathrm{dp}$ of dephosphorylation of target proteins. In this way, a phosphorylated protein contributes only during a time $1/\nu_\mathrm{dp}$ to the cellular response, which we still take to be given by the spatial distribution of the phosphorylation events. In general, it is now a time-dependent quantity, whereas before, we considered the accumulated distribution of all phosphorylation events following a signal. We will assume that phosphorylated proteins do not move. Let us denote the number of phosphorylated proteins resulting from the signal by $N_s$ and those from the background by  $N_\mathrm{bg}$. Let us note again that $N_s$ depends on time and $N_\mathrm{bg}$ does not. To arrive at a single number for the error in estimating the position of Ca$^{2+}$ release, we consider the time point at which $N_s$ is maximal.

We calculate the error of the position estimate by a weighted mean of the error from the phosphorylated proteins resulting from the signal and those from the background. Since, background phosphorylation is independent of phosphorylation in response to the signal, the corresponding variances and thus errors simply add up. The error $\ell_s^2$ associated with the response to the signal is calculated as before. For the error resulting from the background, we assume that the corresponding phosphorylation events are uniformly distributed in the cell, such that $\ell_\mathrm{bg}^2$ is given by the size of the cell. The total error then is
\begin{align}
\ell^2 &=\frac{N_{s,\mathrm{max}}}{N_{s,\mathrm{max}} + N_\mathrm{bg}}\ell_s^2  + \frac{N_\mathrm{bg}}{N_{s,\mathrm{max}} + N_\mathrm{bg}}\ell_\mathrm{bg}^2 .
\end{align} 

\begin{figure}
\includegraphics[width=0.5\textwidth]{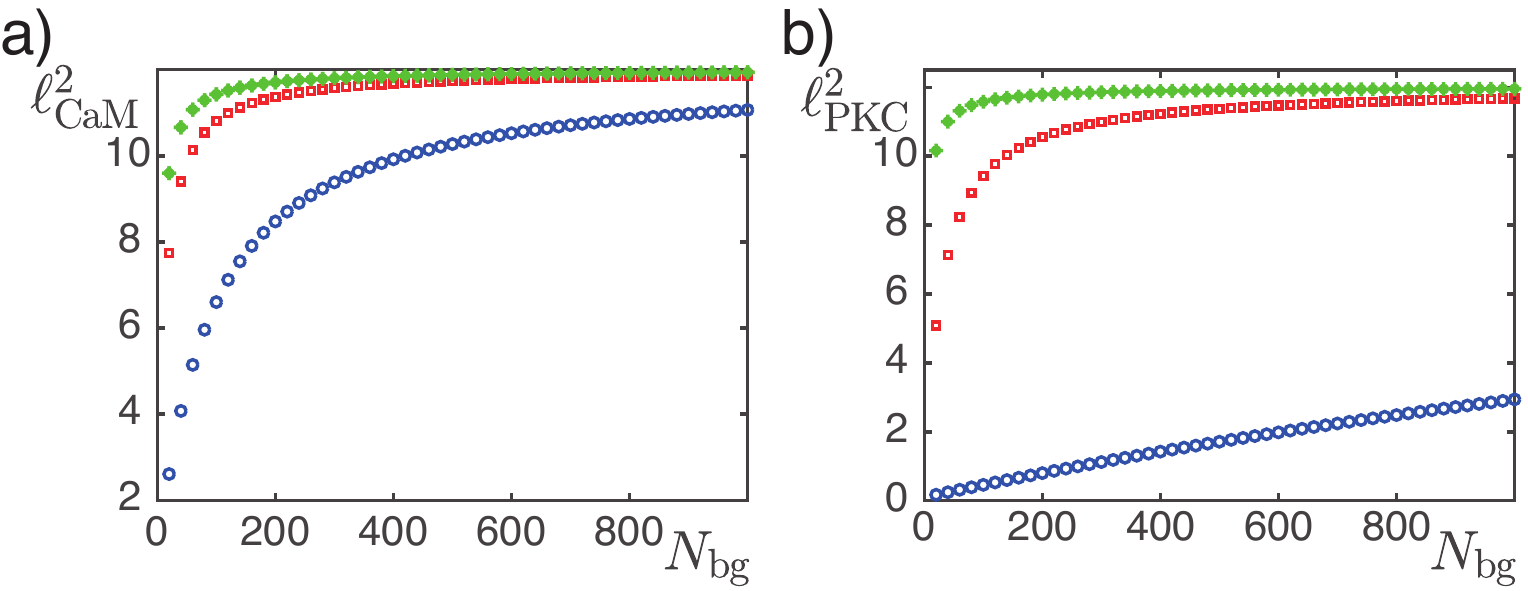}
\caption{\label{fig:background}(color online) Estimation error in presence of background phosphorylation as a function of the number $N_\mathrm{bg}$ of phosphorylated kinases due to background phsophorylation for (a) a cytosolic kinase with $\nu_d=0.1$ (blue circles), 0.9 (red squares), and 1.9 (green stars) and (b) a membrane-binding kinase with $\nu_d=1$ (blue circles), 10 (red squares), and 100 (green stars). Other parameter values are $D_K=0.01$, $\nu_a=0.1$, $\nu_l=10$, $\nu_\mathrm{dp} = 0.001$, $\nu_b=1$, and $\nu_u=0.1$.}
\end{figure}

In Figure~\ref{fig:background}, we present the error for estimating the Ca$^{2+}$ release site for a cytosolic and a membrane-binding kinase in presence of background phosphorylation. As expected, if the background phosphorylation exceeds a certain threshold, the signal is masked and the error equals the size of the cell, such that any information about the site of Ca$^{2+}$ release is lost.

\section{Discussion}

In this work, we have presented a framework for studying cellular responses to localized signals. For concreteness, we have considered the signal to be given by a localized Ca$^{2+}$ influx and the response to be represented by the spatial distribution of phosphorylation events of either a cytosolic or a membrane-binding kinase. We very much simplified the cellular response. For example, we considered direct activation of the cytosolic kinase by binding a Ca$^{2+}$ ion, although often Ca$^{2+}$ activates Calmodulin, which in turn activates the kinase. Furthermore, we assumed that the membrane-binding kinase is activated directly by binding to the membrane, whereas the Protein Kinase C$\alpha$, for example, requires binding to Diacylglycerol (DAG) in the membrane for activation. In spite of these simplifications, we expect our general results to be valid also in more realistic situations. This holds, notably, for the tendency of the estimation error to decrease with increasing Ca$^{2+}$ detachment. Furthermore, membrane-binding kinases should provide a better spatial localization of the signal than cytosolic kinases. Beyond the specific question of how cells can localize external signals, our framework can also be applied in various other situations, in which a stochastic birth process is coupled to diffusion.

\textcolor{black}{We defined the error in determining the position at which the Ca$^{2+}$ ions entered the system. To this end, we considered the variance of the distribution of estimated positions from the generated phosphorylation events. The cell response does not end with the phosphorylation of target proteins. Ultimately, the Ca$^{2+}$ signal should lead to endo- or exocytosis or to a change in the strength of a synapse. However, it is difficult to see how the outcomes of these processes could spatially be more precise than the phosphorylation 'signal' into which the Ca$^{2+}$ signal has been transformed.}

Let us estimate the error in the light of measured parameter values for the Protein Kinase C$\alpha$. The diffusion constant of free cytoplasmic Ca$^{2+}$ is about 500~$\mu$m$^2$/s~\cite{Donahue:1987jx}, whereas cytoplasmic PKC$\alpha$ has a diffusion constant on the order of 10~$\mu$m$^2$/s~\cite{Schaefer:2001fp}. With $\nu_p\approx2$/s, $\nu_d\approx20$/s~\cite{Nalefski:2001wy}, and $\nu_l\approx40/s$~\cite{Smith:1998gg}, the estimation error for a single Ca$^{2+}$ ion is about 7~$\mu$m. This justifies neglecting any boundaries in the lateral direction as a typical cell diameter is 50~$\mu$m. For a Ca$^{2+}$ puff this value decreases further with an increasing number of Ca$^{2+}$ ions. With regards to our assumption that  membrane-binding kinases are immobile, note that the diffusion length on the membrane $\sqrt{D_{K,\textrm{mem}}/\nu_u}\approx0.25~\mu$m. In this estimate, we took the diffusion constant $D_{K,\textrm{mem}}$ of PKC$\alpha$ to be 100 times smaller than its cytoplasmic diffusion constant~\cite{LippincottSchwartz:2001dx} and $\nu\approx7$/s~\cite{Nalefski:2001wy}. Since this length is an order of magnitude smaller than the estimation error, our assumption seems justified.

In future studies of the Protein Kinase C$\alpha$ it should be interesting to consider its dynamics in more detail. In addition to adding the effect of binding DAG to PKC$\alpha$, it might be interesting to consider the full cascade of a signal activating Phospholipase C that breaks Phosphatidylinositol-biphosphate (PIP$_2$) into DAG and Inositol-triphosphate (IP$_3$). The latter diffuses and can open nearby internal Ca$^{2+}$ stores, making the activation of PKC$\alpha$ much more involved than considered here. In addition, PKC$\alpha$ forms clusters on the membrane that affects the lifetime of its activated state~\cite{Bonny:2016em,Swanson:2016jj}. Studying the dynamics of PKC$\alpha$ might yield insights into how its localized activation can help cells to obtain a specific response to external signals even though they get transformed into general purpose second messengers~\cite{Horne:1997ig,Maasch:2000fr}.

\appendix

\section{Full Master equation for the number of phosphorylation events for a membrane-binding kinase}
\label{app:masterEquationPhosphorylationPKC}  

In case, Ca$^{2+}$ can reattach to the kinase after detachment and the kinase can rebind to the membrane after unbinding, the Master equation for the number of phosphorylation events is  
\begin{align}
\label{eq:CndotPKCFull}
\partial_t C_n &= \partial_z^2C_n +\nu_d K_n-\left(\nu_a+\nu_l\right)C_n\\
\label{eq:KndotPKCFull}
\partial_t K_n &=D_K\partial_z^2K_n -\nu_d K_n +\nu_aC_n\\
\label{eq:k0dotPKCFull}
\dot k_0 &= \nu_bK_0(z=0)-\nu_uk_0-k_0\\
\label{eq:kndotPKCFull}
\dot k_n &= \nu_bK_n(z=0)-\nu_uk_n-k_n+k_{n-1}\\
\label{eq:PndotPKCFull}
\dot P_n &=\nu_l\int_0^\infty C_n\;\mathrm{d}z
\end{align}
where Eqs.~(\ref{eq:CndotPKCFull}), (\ref{eq:KndotPKCFull}), and (\ref{eq:PndotPKCFull}) hold for all $n\ge0$, whereas Eq.~(\ref{eq:kndotPKCFull}) holds for all $n\ge1$. In these equations, $C_n$ and $K_n$, respectively, denote the probabilities of having $n$ phosphorylation events with a free Ca$^{2+}$ ion and with it bound to a cytosolic kinase, whereas $k_n$ and $P_n$ denote the respective probabilities with the kinase bound to the membrane and after the Ca$^{2+}$ ion is lost from the system. These equations are complemented by boundary condition for Eqs.~(\ref{eq:CndotPKCFull}) and (\ref{eq:KndotPKCFull}). Explicitly,
\begin{align}
\partial_z\left.C_n\right|_{z=0}&=0\\
\label{eq:bcPKCFull}
-D_K\partial_z\left.K_n\right|_{z=0} &= -\nu_b K_n(z=0) + \nu_u k_n.
\end{align} 
Finally, initially, the Ca$^{2+}$ is located at $z=0$.

In the limit $t\to\infty$, the distribution $P^\infty_n$ of the number of phosphorylation events is given by $P^\infty_n=\nu_l\int_0^\infty \bar{C}_n\;\mathrm{d}z$, where the bar indicates integration of $C_n$ from 0 to $\infty$ with respect to time. For the barred quantities, the equations read
\begin{align}
\partial_z^2\bar{C}_0+\nu_d \bar{K}_0-(\nu_a +\nu_l)\bar{C}_0 &=-\delta(z)  \\
\partial_z^2\bar{C}_n+\nu_d \bar{K}_n-(\nu_a +\nu_l)\bar{C}_n &=0  \\
D_K\partial_z^2\bar{K}_0 - \nu_d\bar{K}_0+  \nu_a  \bar{C}_0  &=0  \\
D_K\partial_z^2\bar{K}_n - \nu_d\bar{K}_n + \nu_a  \bar{C}_n &= 0\\
\nu_b\bar{K}_0(z=0)-\nu_u\bar{k}_0-\bar{k}_0 &= 0\\
\nu_b\bar{K}_n(z=0)-\nu_u\bar{k}_n-\bar{k}_n+\bar{k}_{n-1} &= 0,
\end{align}
where $n\ge1$. Their solution can be written as
\begin{align}
\bar{C}_0 (z)&= \tilde{I}(z) 
 + \hat{C}_0 \left(\lambda_2 \mathrm{e}^{-\lambda_1  z } - \lambda_1 \mathrm{e}^{-\lambda_2  z }\right) \\
\bar{K}_0 (z)&= \nu_a\left[I(z)
- \hat{C}_0\left(\frac{ \lambda_2}{D_K \lambda_1^2 - \nu_d } \mathrm{e}^{-\lambda_1 z } -\frac{ \lambda_1}{D_K \lambda_2^2 - \nu_d } \mathrm{e}^{-\lambda_2 z}\right)\right]\\
\bar{C}_n (z) &= \hat{C}_n\left(\lambda_2 \mathrm{e}^{-\lambda_1  z } -\lambda_1 \mathrm{e}^{-\lambda_2  z } \right)   \\ 
\bar{K}_n(z) &= -\nu_a\hat{C}_n\left(\frac{\lambda_2}{D_K \lambda_1 ^2 - \nu_d } \mathrm{e}^{-\lambda_1  z } +  \frac{\lambda_1}{D_K \lambda_2 ^2 - \nu_d } \mathrm{e}^{-\lambda_2  z }\right)\\
\intertext{with}
I(z) &= \int_0^\infty \frac{\cos(kz) }{\left(D_Kk^2+\nu_d\right)\left(k^2+\nu_a+\nu_l\right)-\nu_a\nu_d} \mathrm{d}k\\
\tilde{I}(z)  &= \int_0^\infty \frac{\left(D_K k^2+\nu_d\right)\cos(kz) }{\left(D_Kk^2+\nu_d\right)\left(k^2+\nu_a+\nu_l\right)-\nu_a\nu_d} \mathrm{d}k.
\end{align}
Using Equations~(\ref{eq:k0dotPKCFull}) and (\ref{eq:kndotPKCFull}) as well as the boundary conditions (\ref{eq:bcPKCFull}), we find
\begin{align}
\hat{C}_0 &= \frac{\nu_bI(0)}{\left(\lambda_1-\lambda_2\right)\left[\left(1+\nu_u\right)A+B\right]}\\
\hat{C}_1 & = -\frac{\nu_b\nu_uA}{(1+\nu_u)A+B}\hat{C}_0\\
\hat{C}_n &= \left[\frac{A+B}{(1+\nu_u)A+B}\right]^{n-1}\hat{C}_1\\
\intertext{with}
A &= D_K\lambda_1\lambda_2\left(\lambda_1+\lambda_2\right)\\
B &= \frac{\nu_b}{\nu_a\nu_d}\left(\lambda_1\lambda_2+\nu_a+\nu_b\right).
\end{align}
This eventually leads to the probability distribution
\begin{align}
P^\infty_0 &= 1 - E\\
P^\infty_n &= \frac{\nu_uA}{(1+\nu_u)A+B}\left[\frac{A+B}{(1+\nu_u)A+B}\right]^{n-1}E\\
\intertext{for $n\ge1$ with}
E&=\frac{\nu_l\nu_b}{(1+\nu_u)A+B}\frac{\lambda_1+\lambda_2}{\lambda_1\lambda_2}I(0).
\end{align}

\acknowledgments
We acknowledge funding through SFB 1027 by Deutsche Forschungsgemeinschaft. \textcolor{black}{The computations were performed at University of Geneva on the Baobab cluster.}

\end{document}